\documentclass[english,prl,lengthcheck,superscriptaddress,floatfix]{revtex4-1}
\usepackage[T1]{fontenc}
\usepackage[utf8]{inputenc}
\usepackage{amsmath}
\usepackage{amssymb}
\usepackage{graphicx}
\usepackage{microtype}

\makeatletter
\usepackage{babel}
\usepackage{color}
\usepackage{babel}
\usepackage{verbatim}
\usepackage{amsmath}
\usepackage{amssymb}
\usepackage{graphicx}
\usepackage[pdfusetitle,colorlinks,linkcolor=blue,citecolor=blue,urlcolor=blue]{hyperref}

\global\hyphenpenalty=100000 
\global\exhyphenpenalty=100000 

\def\equationautorefname~#1\null{Eq.~(#1)\null}

\makeatother

\usepackage{babel}
\begin{document}
\title{Polaritonic Molecular Clock: All-Optical Ultrafast Imaging of Wavepacket
Dynamics without Probe Pulses}
\author{R. E. F. Silva}
\email{ruiefdasilva@gmail.com}

\affiliation{Departamento de Física Teórica de la Materia Condensada and Condensed
Matter Physics Center (IFIMAC), Universidad Autónoma de Madrid, E-28049
Madrid, Spain}
\author{Javier del Pino}
\affiliation{Center for Nanophotonics, AMOLF, Science Park 104, 1098 XG Amsterdam,
The Netherlands}
\author{Francisco J. García-Vidal}
\affiliation{Departamento de Física Teórica de la Materia Condensada and Condensed
Matter Physics Center (IFIMAC), Universidad Autónoma de Madrid, E-28049
Madrid, Spain}
\affiliation{Donostia International Physics Center (DIPC), E-20018 Donostia/San
Sebastián, Spain}
\author{Johannes Feist}
\email{johannes.feist@uam.es}

\affiliation{Departamento de Física Teórica de la Materia Condensada and Condensed
Matter Physics Center (IFIMAC), Universidad Autónoma de Madrid, E-28049
Madrid, Spain}
\begin{abstract}
Conventional approaches to probing ultrafast molecular dynamics rely
on the use of synchronized laser pulses with a well-defined time delay.
Typically, a pump pulse excites a wavepacket in the molecule. A subsequent
probe pulse can then dissociates or ionizes the molecule, and measurement
of the molecular fragments provides information about \emph{where}
the wavepacket was for each time delay. In this work, we propose to
exploit the ultrafast nuclear-position-dependent emission obtained
due to large light-matter coupling in plasmonic nanocavities to image
wavepacket dynamics using only a single pump pulse. We show that the
time-resolved emission from the cavity provides information about
\emph{when} the wavepacket passes a given region in nuclear configuration
space. This approach can image both cavity-modified dynamics on polaritonic
(hybrid light-matter) potentials in the strong light-matter coupling
regime as well as bare-molecule dynamics in the intermediate coupling
regime of large Purcell enhancements, and provides a new route towards
ultrafast molecular spectroscopy with plasmonic nanocavities. 
\end{abstract}
\maketitle

\section*{Introduction}

The interaction of light and matter is one of the most fundamental
ways to unveil the laws of nature and also a very important tool in
the control and manipulation of physical systems. When a confined
light mode and a quantum emitter interact, the timescale for the energy
exchange between both constituents can become faster than their decay
or decoherence times and the system enters the strong coupling regime~\citep{Thompson1992,Weisbuch1992,Lidzey1998}.
In this regime, the excitations of the system become hybrid-light
matter states, so-called polaritons, separated by the vacuum Rabi
splitting $\Omega_{R}$. Due to their relatively large dipole moments
and large exciton binding energies, strong coupling can be achieved
with organic molecules at room temperature down to the few- or even
single-molecule level~\citep{Zengin2015,Chikkaraddy2016,Ojambati2019}.
Strong coupling can lead to significant changes in the behavior of
the coupled system, affecting properties such as the optical response~\citep{Lidzey1998,Vasa2013,Torma2015,Zengin2015,Chikkaraddy2016,Cwik2016,DelPino2018Dynamics,Herrera2018Theory,Singh2018,Ojambati2019},
energy transport~\citep{Coles2014,Orgiu2015,Feist2015,Zhong2017},
chemical reactivity~\citep{Hutchison2012,Galego2015,Herrera2016,Thomas2016,Flick2017Atoms,Munkhbat2018,Fregoni2018,Peters2019,Du2019Remote},
and intersystem crossing~\citep{Munkhbat2018,Stranius2018,Martinez-Martinez2018Polariton}.
However, up to now these setups did not provide direct information
on the molecular dynamics.

A well-known approach to directly probe molecular dynamics is through
the use of ultrashort coherent laser pulses, pioneered in the fields
of femtochemistry~\citep{Zewail2000} and attosecond science~\citep{Krausz2009}.
This allows to observe and control nuclear and electronic dynamics
in atoms and molecules at their natural timescale (fs and sub-fs)
and is a fundamental tool towards a better understanding of chemical
and electronic processes~\citep{Potter1992,Assion1998,Zewail2000,Krausz2009,Kling2006,Corrales2014,Palacios2019}.
In particular, real-time imaging of molecular dynamics can be achieved
in experiments with a pump-probe setup with femtosecond resolution
combined with the measurement of photoelectron spectra~\citep{Assion1998}.
While similar approaches could in principle provide a dynamical picture
of molecules under strong light-matter coupling~\citep{Kowalewski2016Cavity,Vendrell2018Coherent,Triana2019},
common molecular observables (such as dissociation or ionization yields
or photoelectron spectra) are difficult to access in typical experimental
setups, with molecules embedded in a solid-state matrix and confined
within nanoscale cavities~\citep{Zengin2015,Chikkaraddy2016,Ojambati2019}.
Another powerful approach is given by transient absorption spectroscopy,
where the change of the absorption spectrum of a probe pulse is monitored
as a function of time delay after a pump pulse. While this can provide
significant insight about molecular dynamics~\citep{Polli2010},
the interpretation of the spectra is nontrivial due to the competition
between several distinct effects (such as ground-state bleach, stimulated
emission, and excited-state absorption) in the spectrum~\citep{Berera2009},
such that transient absorption spectroscopy only gives an indirect
fingerprint of the molecular dynamics.

\begin{figure}
\includegraphics[width=1\columnwidth]{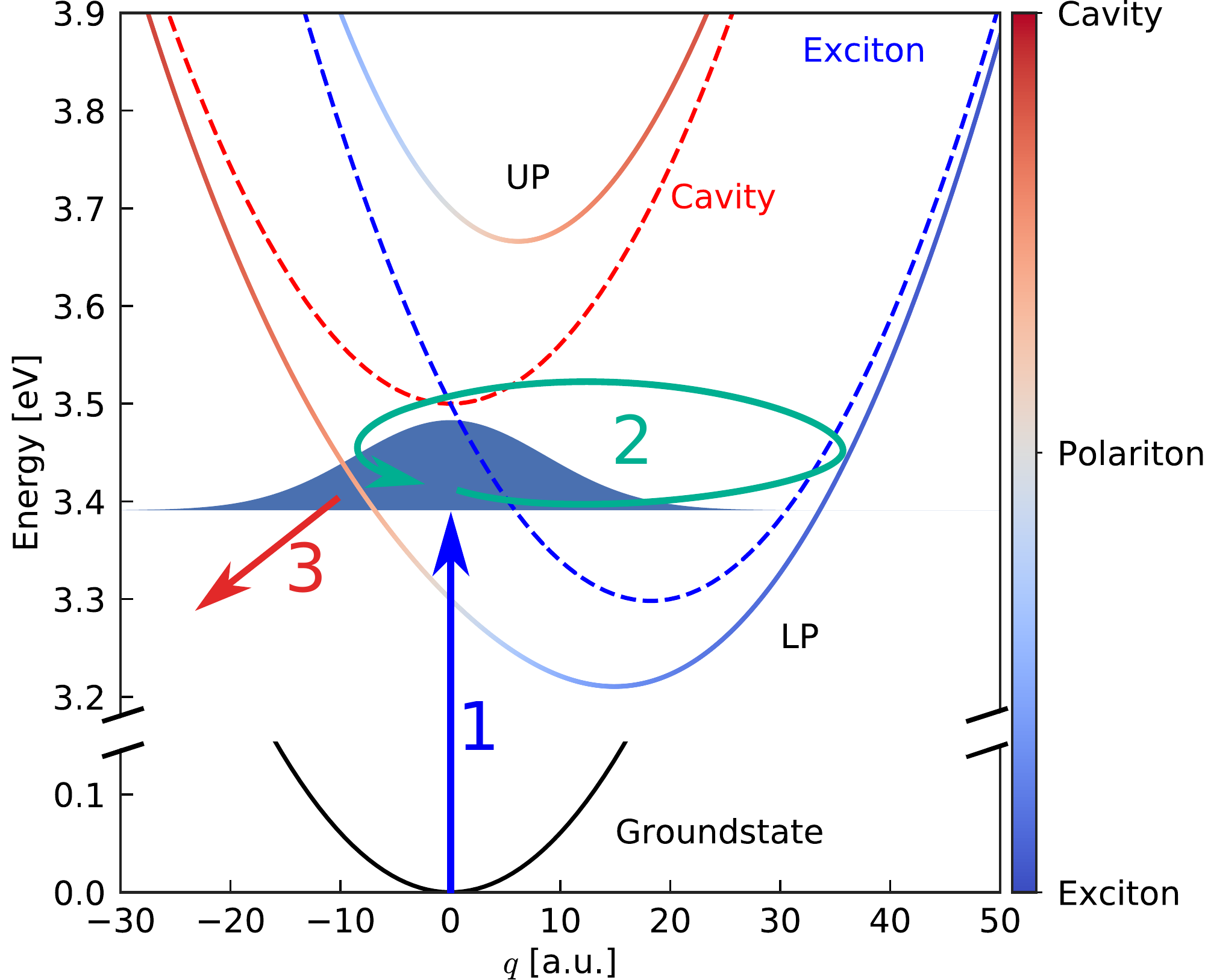} \caption{\label{fig:1} Polaritonic potential energy surfaces (PoPES) in the
single-excitation subspace for a single molecule coupled to a confined
light mode. The red dashed line represents the uncoupled potential
energy surfaces (PES) for a ground-state molecule with a single photon
in the cavity, while the blue dashed line represents the molecular
excited-state PES with no photons present and the black solid line
represents the ground-state PES of the molecule. The solid blue/gray/red
curves are the lower and upper polariton PES, with the color encoding
the excitonic/polaritonic/photonic character as a function of nuclear
position $q$. The filled blue curve represents the vibrational ground-state
wavefunction of the electronic ground-state PES\@. The arrows represent
the excitation by the laser pulse (1), oscillatory motion of the excited
vibrational wavepacket (2) and radiative emission (3).}
\end{figure}

In this article, we demonstrate that the ultrafast emission induced
by strong coupling to plasmonic modes can be used to monitor molecular
wavepacket dynamics by measuring the time-resolved light emission
of the system after excitation by an ultrashort laser pulse, without
the need of a synchronized probe pulse. Our approach exploits the
fact that the light-matter hybridization in a molecule is nuclear-position-dependent.
Consequently, efficient emission only occurs in regions where the
polaritonic potential energy surface (PoPES)~\citep{Galego2015,Feist2018}
on which the nuclear wavepacket moves possesses a significant contribution
of the cavity mode, as sketched in \autoref{fig:1}. Additionally,
due to the very low lifetime (or equivalently, low quality factor)
of typical plasmonic nanocavity modes on the order of femtoseconds,
emission from the cavity also becomes an ultrafast process. Instead
of using a probe pulse to learn \emph{where} the nuclear wavepacket
is at a given time delay, we thus use the nuclear-position-dependent
emission to learn \emph{when} the wavepacket passes a given spatial
region. Tracking the time-dependent emission from the cavity then
gives direct information about the nuclear dynamics by effectively
clocking the time it takes the wavepacket to perform a roundtrip in
the PoPES through an all-optical measurement. We note that a variety
of experimental techniques allow the measurement of time-dependent
light pulses with few-femtosecond resolution, e.g., intensity cross-correlation~\citep{Nicholson1999},
SPIDER~\citep{Iaconis1998}, FROG~\citep{Trebino1997} or d-scan~\citep{Miranda2012}.

\section*{Results}

We first illustrate these ideas using a minimal model system: A single-mode
nanocavity containing a molecule with two electronic states and a
single vibrational degree of freedom, which for simplicity we approximate
as a harmonic oscillator (with displacement between the ground and
excited state due to exciton-phonon coupling). Our model is then equivalent
to the Holstein-Jaynes-Cummings model that has been widely used in
the literature to model strongly coupled organic molecules~\citep{Michetti2009,Kirton2013,Spano2015,Herrera2016},
with the main difference that we explicitly treat cavity losses and
driving by an ultrashort (few-fs) laser pulse, and monitor the time-dependent
emission. While this is a strongly reduced model which allows for
a straightforward interpretation, we will later show that the results
we observe are also obtained in realistic simulations of molecules
with a plethora of vibrational modes leading to rapid dephasing~\citep{DelPino2018Dynamics}.
The system is described by the Hamiltonian (setting $\hbar=1$) 
\begin{multline}
H(t)=\omega_{e}\sigma^{+}\sigma^{-}+\frac{p^{2}}{2}+\omega_{v}^{2}\frac{q^{2}}{2}-\lambda_{v}\sqrt{2\omega_{v}}\sigma^{+}\sigma^{-}q\\
+\omega_{c}a^{\dagger}a+\frac{\Omega_{R}}{2}(a^{\dagger}\sigma^{-}+a\sigma^{+})+\mu_{c}E(t)(a^{\dagger}+a),\label{eq:Ham}
\end{multline}
where $\sigma^{+}$ ($\sigma^{-}$) is the raising (lowering) operator
for the electronic state with excitation energy $\omega_{e}=3.5~$eV,
while $p$ and $q$ are the mass-weighted nuclear momentum and position
operators for the vibrational mode with frequency $\omega_{v}=0.182~$eV
and exciton-phonon coupling strength $\lambda_{v}=0.192~$eV (with
these parameters we reproduce the properties of the anthracene molecule,
see Methods for further details). The cavity is described through
the photon annihilation (creation) operators $a$ ($a^{\dagger}$),
with a photon energy chosen on resonance with the exciton, $\omega_{c}=\omega_{e}$.
In addition to the coherent dynamics described by the Hamiltonian,
the cavity mode decays with rate $\gamma_{c}=0.1~$eV, described by
a standard Lindblad decay operator (see Methods for details). The
photon-exciton coupling is described through the Rabi splitting at
resonance, $\Omega_{R}=2\vec{E}_{1\mathrm{ph}}(\vec{r}_{m})\cdot\vec{\mu}_{eg}$,
where $\vec{E}_{1\mathrm{ph}}(\vec{r}_{m})$ is the quantized mode
field of the cavity at the molecular position, and $\vec{\mu}_{eg}$
is the transition dipole moment of the molecule (in principle, this
is $q$-dependent, but is taken constant here for simplicity). Finally,
the cavity mode is coupled through its effective dipole moment $\mu_{c}$
to an external (classical) laser pulse $E(t)=E_{0}\cos(\omega_{L}t)\exp(-\sigma_{L}^{2}t^{2}/2)$,
with central frequency $\omega_{L}$, spectral bandwidth $\sigma_{L}$
and a corresponding duration of $\approx1.67/\sigma_{L}$ (FWHM of
intensity). We note that since the cavity mode is driven by the external
field, the effective pulse felt by the molecule (in particular in
the weak-coupling limit) is slightly distorted and not just given
by $E(t)$.

\begin{figure}
\includegraphics[width=1\columnwidth]{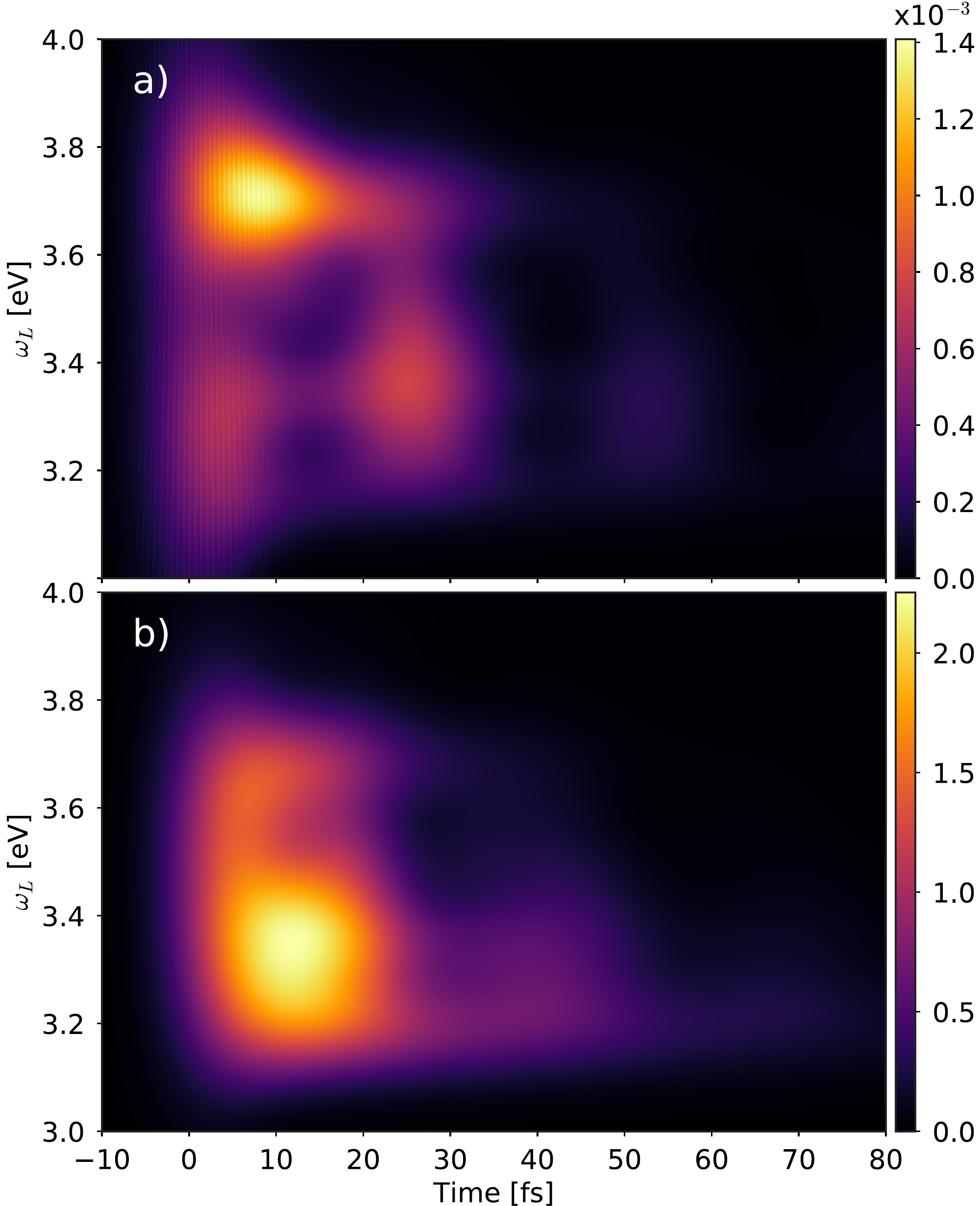} \caption{\label{fig:2}Mean values of the time-dependent radiative emission,
$E_{R}$, (a) and $\langle\sigma^{+}\sigma^{-}\rangle$ (b) for different
values of $\omega_{L}$ and for $\Omega_{R}=0.4~$eV. For all calculations,
$E_{0}=2.1\times10^{-7}\,\mathrm{a.u.}$ and $\sigma_{L}=0.1$~eV.}
\end{figure}

\begin{figure}
\includegraphics[width=1\columnwidth]{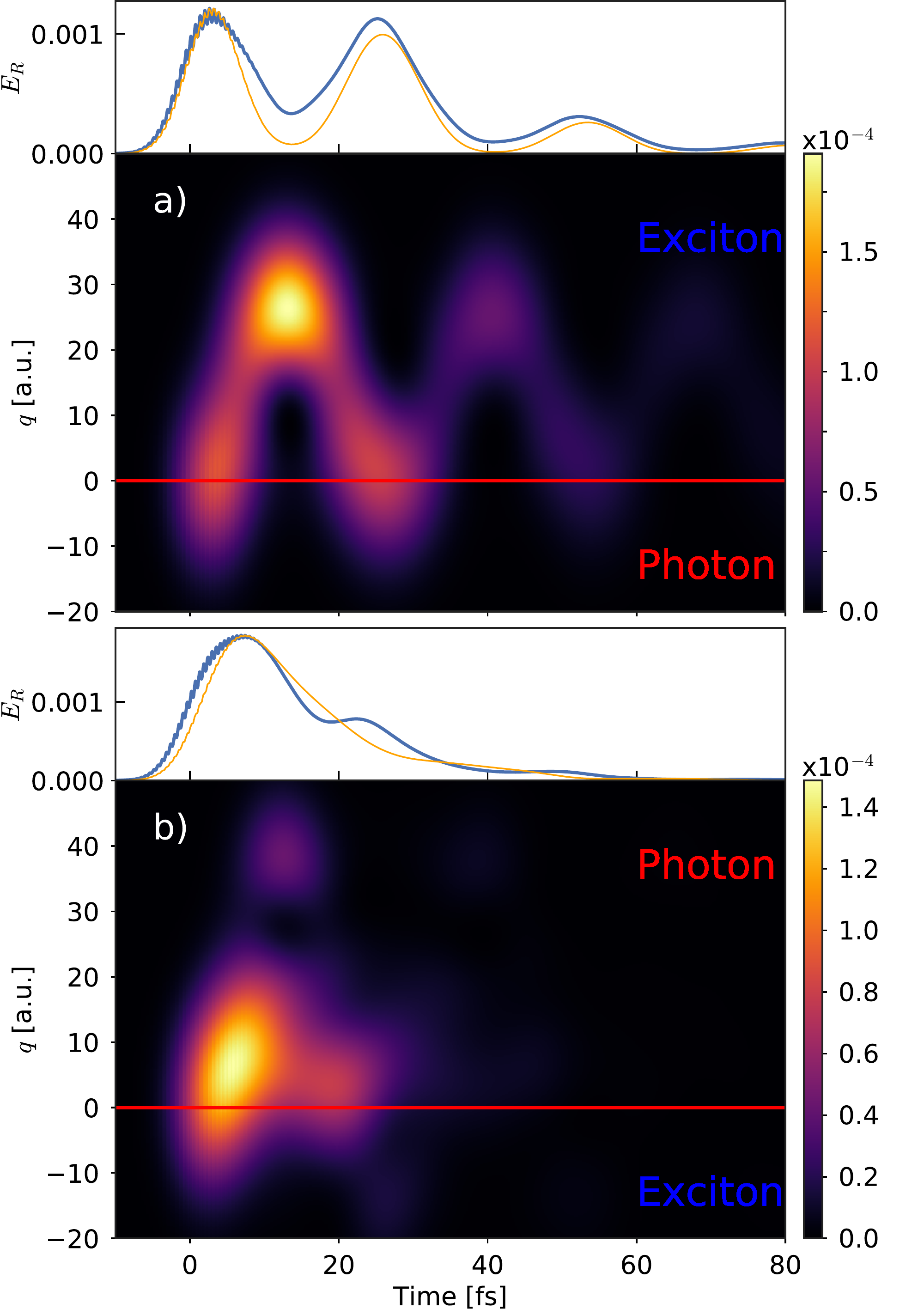} \caption{\label{fig:3} Probability density of the vibrational wavepackets
in the single-excitation subspace for two different laser frequencies:
(a) $\omega_{L}=3.3~$eV and (b) $\omega_{L}=3.7~$eV. The red line
at $q=0$ indicates the border where the polariton switches from photonic
to excitonic character for the (a) lower and (b) upper polariton.
The upper panels in each subfigure show the time-dependent emission
from the cavity (thick blue lines) and the (scaled) probability of
the nuclear wavepacket on the photonic side, given by $q<0$ for (a)
and $q>0$ for (b) (orange lines). For both calculations, $E_{0}=2.1\times10^{-7}\,\mathrm{a.u.}$
and $\sigma_{L}=0.15~$eV.}
\end{figure}

We start by analyzing the system response in the strong-coupling regime
($\Omega_{R}=0.4~$eV) after excitation by an ultrashort laser pulse
with $\sigma_{L}=0.1~$eV, while scanning the laser frequency $\omega_{L}$.
For $\sigma_{L}=0.1$~eV, the duration of the pulse is $\approx11$~fs.
The laser intensity is chosen small enough to remain in the single-excitation
subspace (i.e., within linear response). The instantaneous radiative
emission rate from the cavity is given by $E_{R}=\gamma_{c,r}\langle a^{\dagger}a\rangle$,
where $\gamma_{c,r}$ is the radiative decay rate of the cavity excitations.
As it corresponds to a constant (system-dependent) factor, we set
it to unity in the figures shown in the following. Estimates of the
achievable photon yields in realistic systems are given in the discussion
section. In \autoref{fig:2}, the time-dependent radiative emission
intensity $E_{R}$, and the exciton population, $\langle\sigma^{+}\sigma^{-}\rangle$,
are shown. We observe that when the laser pulse is resonant with the
lower polariton region, i.e., for $\omega_{L}$ between $3.2$ and
$3.5$ eV, the cavity emission is modulated in time with a period
of around $26~$fs, while no such oscillation is observed when the
upper polariton branch is excited for $\omega_{L}$ between 3.5 and
3.8 eV. This behavior can be understood with the help of the PoPES,
shown in \autoref{fig:1}. They are obtained by treating nuclear
motion within the Born-Oppenheimer approximation, i.e., with $q$
treated as an adiabatic parameter (see Methods for details). Within
the Franck-Condon approximation, short-pulse excitation creates a
copy of the vibrational ground-state (centered at $q=0$) on the relevant
polaritonic PES\@. This vibrational wavepacket will then evolve on
the potential surface, performing oscillatory motion, with the character
of the wavepacket also oscillating between photon-dominated and exciton-dominated
depending on nuclear position. However, as radiative emission of the
cavity mode is orders of magnitude faster than from the bare molecule
(typically, femtoseconds compared to nanoseconds), efficient emission
is only possible in regions where the relevant PoPES has a significant
photon contribution. Focusing first on the lower polariton, this condition
is fulfilled for $q<0$ for the parameters chosen here, explaining
the observed temporal modulation of the emission intensity, which
effectively corresponds to clocking of the nuclear wavepacket motion.
Furthermore, the period of this motion is determined by the curvature
of the lower polariton PoPES, which is different to the bare-molecule
oscillation period $T_{v}\approx22.7~$fs. Fitting the lower polariton
curve to a harmonic oscillator for the current parameters gives an
oscillation period of $25.9~$fs, in excellent agreement with the
observed modulation frequency of $26~$fs. The temporal emission modulation
thus also provides a direct fingerprint of the strong-coupling induced
modifications of molecular structure. On the other hand, excitation
to the upper polariton creates a wavepacket that spends most of its
time in the region with efficient emission ($q>0$ for the upper PoPES),
such that no clear oscillation between photonic and excitonic character,
and thus no modulation in the emission intensity, is observed.

Up to now, we have confirmed that molecular dynamics imprints its
fingerprint in the time-dependent radiative emission of the cavity.
We now demonstrate that the time-resolved emission intensity indeed
provides a direct quantitative probe of the nuclear wavepacket dynamics.
In \autoref{fig:3}, we show the nuclear probability density $|\psi(q)|^{2}$
in the single-excitation subspace under resonant excitation of the
lower polariton, \autoref{fig:3}(a), and upper polariton, \autoref{fig:3}(b),
respectively. For case (a), the wavepacket starts periodic motion
around the minimum of the lower polariton curve, $q_{\mathrm{min}}\approx15\ \mathrm{a.u.}$,
after the initial excitation at $t\approx0$. In the upper panel,
we show $E_{R}$ and the probability to find the nuclei at $q\leq0$,
given by $\int_{-\infty}^{0}|\psi(q)|^{2}\mathrm{d}q$. The observed
good agreement demonstrates that it is possible to track the position
of the nuclear wavepacket in time through the emission from the cavity.
The similarly good agreement found in \autoref{fig:3}(b), with
the integral this case performed for $q\geq0$ corresponding to excitation
of the upper polariton branch reinforces this notion. We again observe
that less pronounced oscillation is observed for excitation of the
UP branch. We also note that for case (b), there is a small contribution
of the lower polariton to the excitation (since this is energetically
still allowed), explaining the slightly worse agreement between the
full calculation and the simplified approximation based on integrating
the nuclear probability density.

\begin{figure}
\includegraphics[width=1\columnwidth]{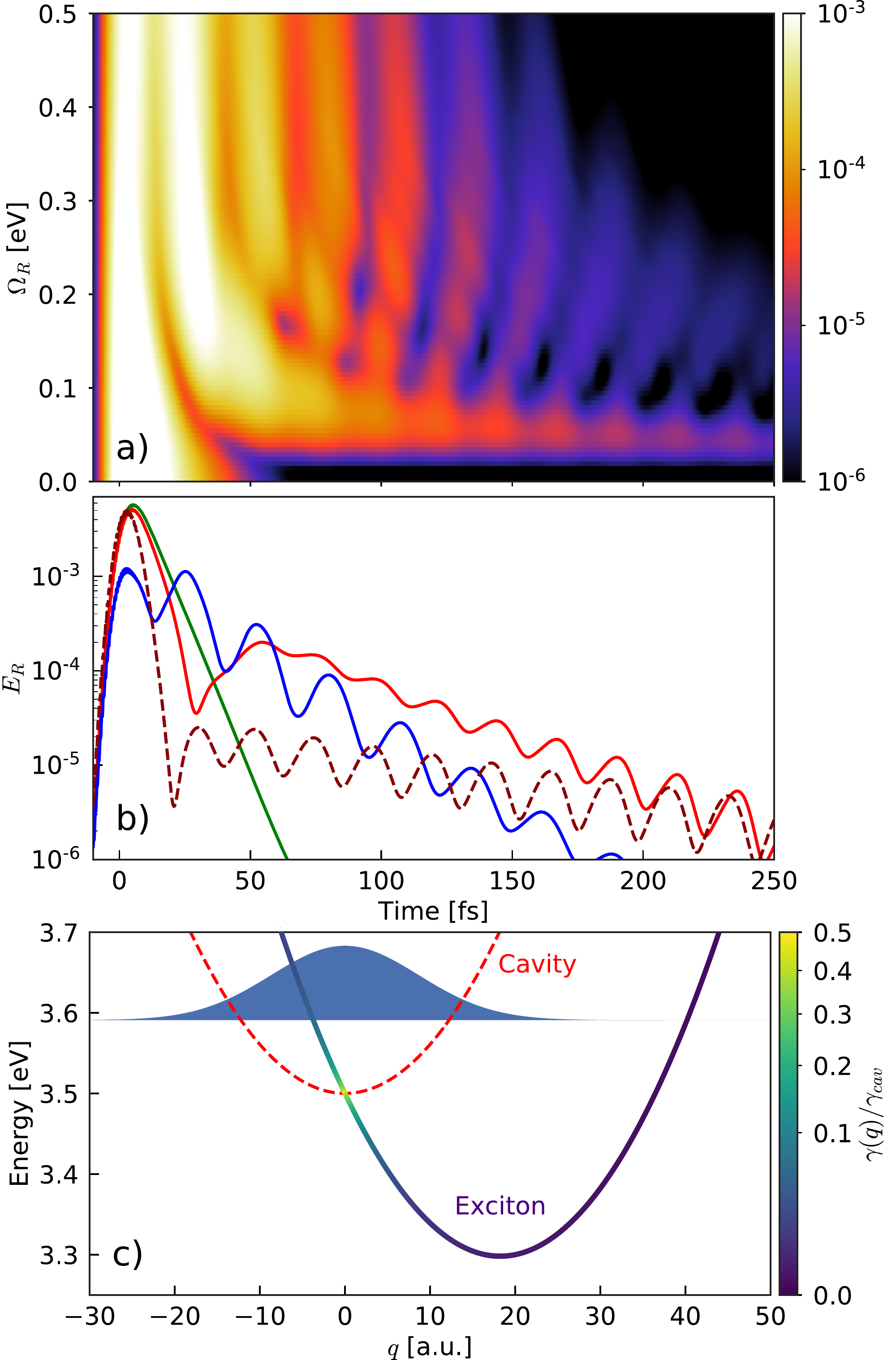} \caption{\label{fig:4}(a) Time-dependent radiative emission from the cavity,
$E_{R}$, for different values of $\Omega_{R}$ and laser frequency
resonant with the bare lower polariton energy, $\omega_{L}=\omega_{e}-\Omega_{R}/2$.
For all calculations, $E_{0}=2.1\times10^{-7}\,\mathrm{a.u.}$ and
$\sigma_{L}=0.15~$eV. (b) Same as in (a) for three different values
of $\Omega_{R}=0.01,\ 0.07$ and $0.4$ $\mathrm{eV}$(green, red
and blue lines respectively). The dashed dark red line represents
the case for $\Omega_{R}=0.07\ \mathrm{eV}$ and a cavity with larger
decay rate, $\gamma_{c}=0.3\ \mathrm{eV}$. (c) Potential energy surfaces
in the weak-coupling regime with large Purcell enhancement of the
emission. The red dashed line represents the PES of the molecule in
its ground-state with a photon in the cavity. The blue-yellow solid
line represents the molecular excited-state PES with no photons present,
with the position-dependent (Purcell-enhanced) decay rate encoded
in the purple/yellow color scale. The filled blue curve represents
the vibrational ground-state wavefunction of the electronic ground-state
PES\@.}
\end{figure}

We next investigate the dependence of the effects discussed above
on the Rabi splitting $\Omega_{R}$, focusing in particular on the
case of smaller $\Omega_{R}$, which would correspond to the weak-coupling
regime. The corresponding time-resolved radiative emission $E_{R}$
is shown in \autoref{fig:4}(a) on a logarithmic scale. Since we
have observed the lower polariton branch to display more interesting
dynamics, the central laser frequency is chosen such that the lower
polariton branch is excited for each Rabi frequency, i.e., $\omega_{L}=\omega_{e}-\Omega_{R}/2$.
Several regimes can be clearly distinguished: For small coupling,
$\Omega_{R}\lesssim0.03~$eV, the molecules barely participate in
the dynamics and the response is dominated by the excitation and subsequent
ringdown (with time constant $\tau_{c}=\hbar/\gamma_{c}\approx6.6$~fs)
of the bare cavity mode (green line in \autoref{fig:4}(b)). In
contrast, within the strong-coupling regime, $\Omega_{R}\gtrsim0.10~$eV,
the previously discussed oscillations can be seen, with the modulation
frequency increasing concomitantly with $\Omega_{R}$ due to the increasingly
large modification of the polaritonic PES and thus the nuclear oscillation
period (blue line in \autoref{fig:4}(b)). For intermediate values
of $\Omega_{R}$, a slightly different behavior is observed: emission
occurs over relatively long times, but is again modulated over time,
with a period of around $23~$fs, in good agreement with the bare-molecule
vibrational period, $T_{v}\approx22.7~$fs. This can be understood
by examining the molecular PES in the case of weak coupling, as shown
in \autoref{fig:4}(c). In that case, the potential energy surfaces
are almost unmodified and the initial laser pulse only excites the
cavity mode, but the relatively large coupling is sufficient to allow
efficient energy transfer to the molecule (exactly in the Franck-Condon
region) within the lifetime of the cavity mode, such that the emission
is not fully dominated by the cavity response. The molecular wavepacket
then again oscillates, now within the bare molecular excited-state
PES\@. However, for nuclear configurations where the molecular exciton
and the cavity mode are resonant (within the cavity bandwidth), the
molecular radiative decay is enhanced strongly through the Purcell
effect, leading to ultrafast emission exactly when the nuclear wavepacket
crosses the resonant configuration ($q\approx0$ for the parameters
considered here). In the intermediate coupling regime, it is important
to point out that the oscillations will be more clear when the cavity
has an ultrafast decay. This can be seen when comparing the radiative
emission for two different decay rates, $\gamma_{c}=0.1$ and $0.3\ \mathrm{eV}$
(solid red and dashed dark red lines in \autoref{fig:4}(b)), where
the oscillations are more prominent for more lossy cavities. We note
that the more relaxed requirements for $\Omega_{R}$ in this intermediate
regime should make it more easily accessible in controlled experimental
setups~\citep{Ojambati2019}.

\begin{figure}
\includegraphics[width=1\columnwidth]{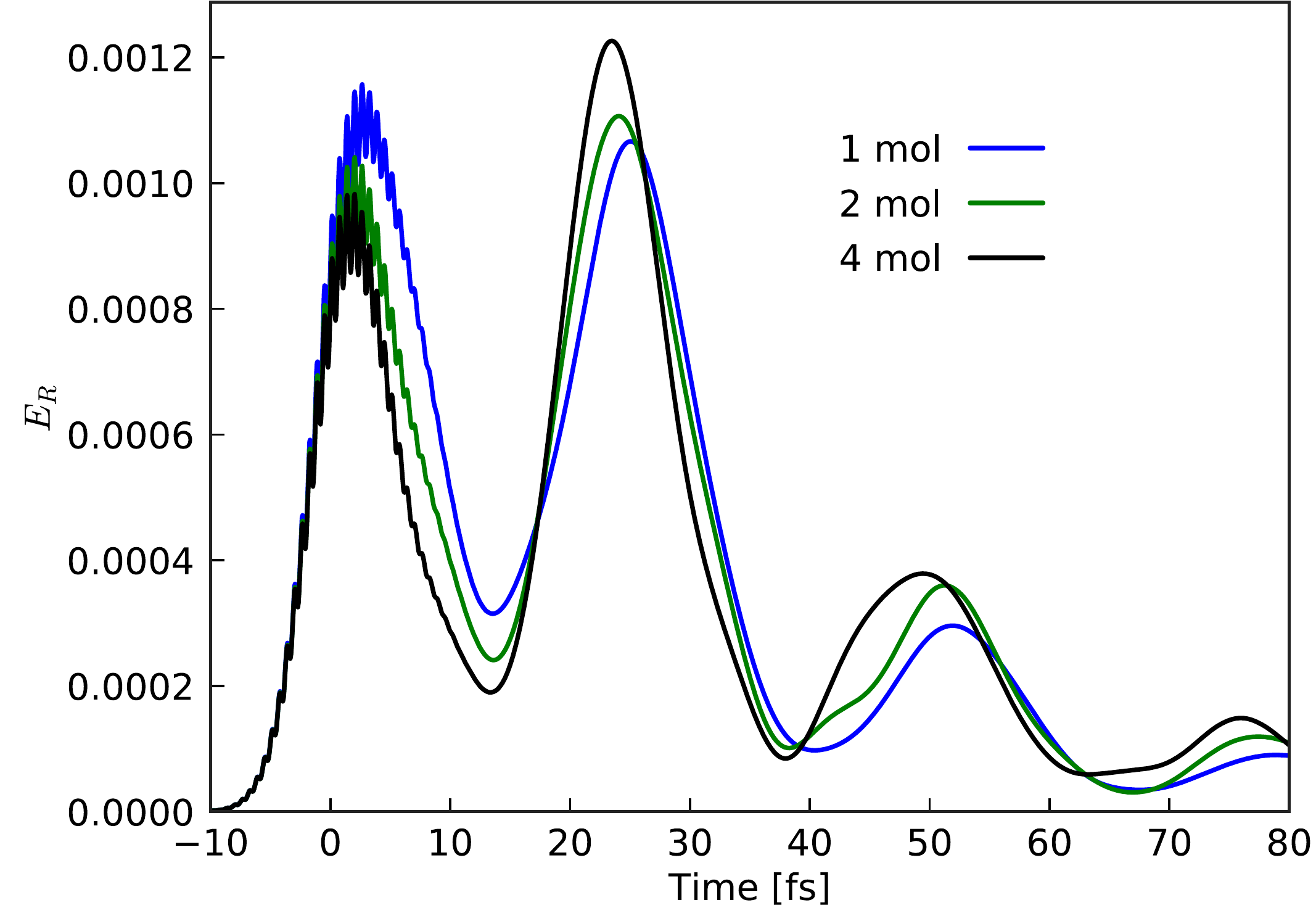} \caption{\label{fig:9}Comparison of the radiative emission for different numbers
of molecules coherently coupled to the same cavity mode while keeping
the total Rabi splitting $\Omega_{R}\propto\sqrt{N}$ fixed. The blue,
green, and black lines correspond to 1, 2, and 4 molecules, respectively.
The frequency of the laser pulse is $\omega_{L}=3.3$$\mathrm{eV}$
and its bandwidth is $\sigma_{L}=0.15~$eV.}
\end{figure}

Up to now, we have focused on the case of a single molecule under
strong coupling. While this serves to highlight the principal properties
of the setup, it is still extremely challenging to achieve in experiment.
On the other hand, collective strong coupling can yield significant
Rabi splittings in available plasmonic nanocavities even for small
numbers of molecules (e.g., $200$~meV for three or four molecules~\citep{Chikkaraddy2016}).
In this situation, several molecules are coherently coupled to the
same photonic mode, with the collective Rabi splitting scaling as
$\sqrt{N}$. In \autoref{fig:9}, we demonstrate that the polaritonic
molecular clock also works in this situation. We plot the time-resolved
radiative emission for $N=1,2$, and $4$ molecules while keeping
the collective Rabi splitting fixed at $\Omega_{R}=0.4$~eV for easier
comparison. This shows that the coherent wavepacket motion of multiple
molecules moving on a collective polaritonic potential energy surface
can be accessed directly with our setup. We note that this is in strong
contrast to standard pump-probe techniques, where only single-molecule
observables are typically interrogated. In contrast, the PoPES in
the case of collective strong coupling describe nuclear motion of
the polaritonic ``supermolecule''~\citep{Galego2016,Galego2017}
and depend on all molecular coordinates. In the Supplementary Information
we use the time-dependent variational matrix product state (TDVMPS)
approach~\citep{DelPino2018Dynamics,Schroder2019} to show that this
approach works even when taking into account all vibrational degrees
of freedom and the associated dephasing. In particular, the effect
of dephasing is not significantly stronger in the many-molecule case
than for a single molecule. Consequently, the proposed setup could
provide a route to directly probe multi-molecule coherent nuclear
wavepacket motion.

\begin{figure}
\includegraphics[width=1\columnwidth]{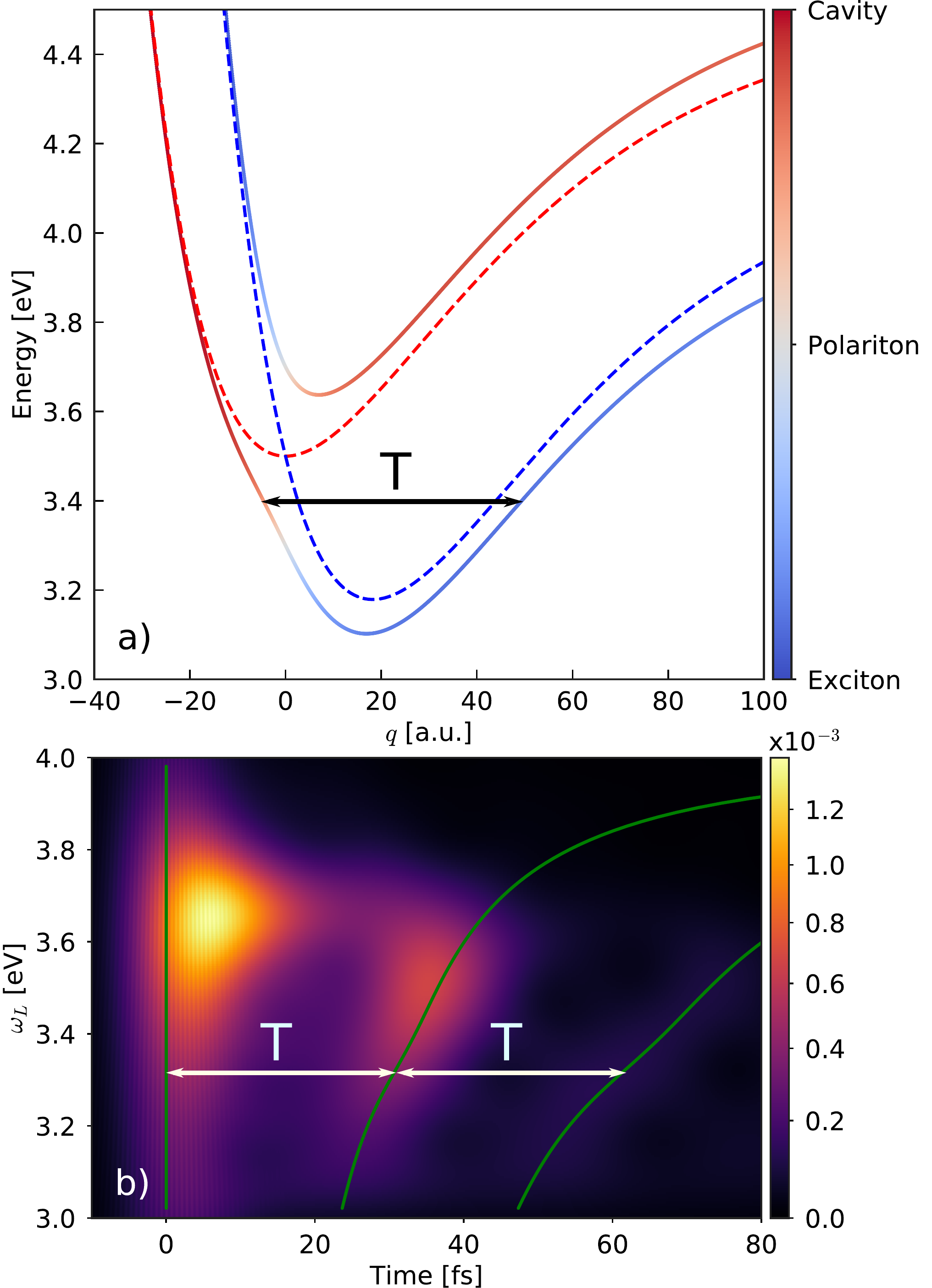} \caption{\label{fig:morse} (a) PoPES for a molecule where the ground and excited
state are described by displaced Morse potentials. The colors and
line styles have the same meaning as in \autoref{fig:1}. (b) Time-dependent
radiative emission $E_{R}$ for different values of $\omega_{L}$
and for $\Omega_{R}=0.4~$eV. For all calculations, $E_{0}=2.1\times10^{-7}\,\mathrm{a.u.}$
and $\sigma_{L}=0.1$~eV. The three green lines are calculated as
$t=nT(\omega_{L})$, with $n=0,1,2$ (see main text).}
\end{figure}

We next demonstrate that our approach is not restricted to displaced
harmonic oscillator models and also gives direct insight into molecular
dynamics in more complex potentials. To that end, we treat a molecule
described by displaced Morse potentials, corresponding to anharmonic
oscillators. The molecular Hamiltonian is given by 
\begin{gather}
H_{m}=\frac{p^{2}}{2}+V_{g}\left(q\right)\sigma^{-}\sigma^{+}+V_{e}\left(q\right)\sigma^{+}\sigma^{-},\\
V_{j}(q)=\delta_{j}+D\left(1-e^{-a(q-q_{j})}\right)^{2},
\end{gather}
where $j\in\{g,e\}$, with parameters $D=1.0$~eV, $a=0.025$~a.u.,
$q_{g}=0$, $q_{e}=18.2$~a.u., $\delta_{g}=0$, and $\delta_{e}=3.18$~eV.
\autoref{fig:morse}(a) shows the uncoupled PES and the corresponding
PoPES under strong coupling, while \autoref{fig:morse}(b) shows
the time-dependent radiative emission as a function of the driving
laser frequency. In contrast with the simple displaced-harmonic-oscillator
model treated before, the oscillation period of the time-dependent
radiative emission now depends on the laser frequency. The insight
this provides into the polaritonic PES becomes clear by comparing
the peak times of the cavity emission with the energy-dependent classical
oscillation period within the lower PoPES, $T(\omega_{L})=2\int_{q_{\mathrm{min}}}^{q_{\mathrm{max}}}\mathrm{d}q\left[\frac{2}{M}(E_{gs}+\omega_{L}-V_{LP}(q))\right]^{-1/2}$,
where $E_{gs}$ is the ground-state energy. The green lines in \autoref{fig:morse}(b)
show that the peak emission happens exactly at $t=nT(\omega_{L})$,
with $n=0,1,2,\ldots$, demonstrating that the polaritonic molecular
clock captures the nuclear wave-packet motion accurately and provides
a direct picture of the dynamics also in non-harmonic potentials.
In the Supplementary Information, we furthermore show that our scheme
could also be used to study photodissocation dynamics within the weak
coupling regime for a model molecule similar to methyl iodide~\citep{Corrales2014}.

We next discuss the requirements that must be fulfilled for the phenomena
described above to be observed. First, the molecule needs to have
sufficiently strong exciton-phonon coupling (i.e., a sufficiently
large change in the $q$-dependent excitation frequency) to lead to
significant spatial modulation of the cavity and exciton components
of the PoPES\@. Furthermore, the slope of the (polaritonic) PES in
the Franck-Condon region has to be large enough for the nuclear wavepacket
to leave the initial position before it has time to decay completely
(although this problem could be mitigated by, e.g., choosing the cavity
to be resonant in another region of nuclear configuration space instead
of at the equilibrium configuration). For the Holstein-type molecular
model studied here, these conditions are satisfied if $\lambda_{v}$
is comparable to the vibrational frequency $\omega_{v}$, and both
are comparable to the cavity decay rate $\gamma_{c}$. These properties
are fulfilled for several organic molecules that have been used in
strong-coupling experiments, such as anthracene~\citep{Kena-Cohen2010}
or the rylene dye {[}N,N0-Bis(2,6-diisopropylphenyl)-1,7- and -1,6-bis(2,6-diisopropylphenoxy)-
perylene-3,4:9,10- tetracarboximide{]}~\citep{Ramezani2017Plasmon}.
Additionally, in order to be able to observe coherent wavepacket motion,
internal vibrational relaxation and dephasing, which typically occurs
on the scale of tens to hundreds of femtoseconds in solid-state environments,
must be slow enough compared to the dynamics of interest. In the Supplementary
Information, we demonstrate that this is the case for the anthracene
molecule by comparing the Holstein-Jaynes-Cummings-model calculation
with large-scale quantum dynamics simulations including all vibrational
modes of the molecule, performed using the time-dependent variational
matrix product state (TDVMPS) approach~\citep{DelPino2018Dynamics,Schroder2019}.

To summarize, we have proposed a novel scheme to probe and image molecular
dynamics by measuring the time-dependent radiative emission obtained
after short-pulse excitation of a system containing few molecules
and a nanocavity with large light-matter coupling, close to or within
the strong-coupling regime. We show that this approach enables to
retrieve a direct mapping of nuclear wavepacket motion in the time
domain, also in the few-molecule case, where this scheme provides
a direct fingerprint of coherent multi-molecular nuclear dynamics.
In the strong-coupling regime, this gives access to the cavity-modified
molecular dynamics occurring on the PoPES, while in the weak-coupling
regime, it allows probing of the bare-molecule excited-state dynamics.
By exploiting the ultrafast emission dynamics in typical highly lossy
plasmonic nanocavity, we obtain the time-resolved dynamics without
the need for a pump-probe setup with synchronized femtosecond pulses.
In addition, in contrast to the common approaches of femtochemistry,
our proposed scheme does not require direct access to molecular observables
such as photoelectron spectra or fragmentation yields, which are difficult
to obtain for typical experimental geometries. Instead, it only relies
on optical access to the nanocavity mode. Additionally, the scheme
only depends on the properties of the first few electronic states
of the molecules, and is not affected by, e.g., the multitude of ionization
channels that have to be taken into account in photoionization~\citep{Palacios2019}.
Since only a single excitation is imparted to the molecules and the
dynamics are probed through the photons emitted upon relaxation to
the ground-state, the molecules are left intact after the pulse. At
the same time, this implies that the absolute photon numbers to be
measured are small. This could be mitigated by using high-repetition-rate
sources (readily available for the low laser intensities required),
as well as collecting the response from an array of identical nanocavities,
taking advantage of highly reproducible setups available nowadays,
e.g., through DNA origami~\citep{Acuna2012,Ojambati2019}. Finally,
we mention that while the cavity decay rate $\gamma_{c}$ in a plasmonic
cavity is typically large and leads to few-femtosecond lifetimes as
required for the discussed approach, this rate is often dominated
by nonradiative contributions that do not lead to far-field emission.
However, fortunately the same plasmonic nanocavity architectures that
provide the current largest coupling strengths, such as nanoparticle-on-mirror
geometries, also provide a significant radiative quantum yield of
close to 50 percent~\citep{Akselrod2014Probing,Kongsuwan2018,Baumberg2019}.

\section*{Methods}

The time dynamics is described by the following Lindblad master equation
\begin{align}
\dot{\rho}(t) & =-i[H(t),\rho(t)]+\gamma_{c}\mathcal{L}_{a}[\rho(t)],\label{eq:lindblad}
\end{align}
where $\mathcal{L}_{a}[\rho(t)]=a\rho(t)a^{\dagger}-\frac{1}{2}[\rho(t)a^{\dagger}a+a^{\dagger}a\rho(t)]$
is a standard Lindblad decay term modelling the incoherent decay of
the cavity mode due to material and radiative losses. The polaritonic
potential energy surfaces used for the interpretation and analysis
of the results are obtained by diagonalizing the (undriven) Hamiltonian
within the Born-Oppenheimer approximation, i.e., diagonalizing $H(t)-p^{2}/2$
for $E_{0}=0$ and fixed $q$~\citep{Galego2015}. In \autoref{fig:1}
we show the PoPES within the single-excitation subspace, spanned by
the uncoupled states $|e,0\rangle$ and $|g,1\rangle$, where $|g\rangle$
($|e\rangle$) is the electronic ground (excited) state and $|n=0,1,\ldots\rangle$
is the cavity mode Fock state with $n$ photons. The Hamiltonian in
this subspace can be written as 
\begin{equation}
H_{BO}(q)=\begin{pmatrix}\omega_{c}+\frac{\omega_{v}^{2}q^{2}}{2} & \Omega_{R}/2\\
\Omega_{R}/2 & \omega_{e}+\frac{\omega_{v}^{2}q^{2}}{2}-\lambda_{v}\sqrt{2\omega_{v}}q
\end{pmatrix},
\end{equation}
and diagonalizing it gives the PoPES plotted in \autoref{fig:1}
and \autoref{fig:4}(b).

The parameter values chosen for modelling the molecule were based
on ab-initio calculations for the anthracene molecule at the TDA-B3LYP
level of theory using Gaussian 09~\citep{Gaussian09E01}. Fitting
the PES obtained in these calculations to a displaced harmonic oscillator
model using the Duschinsky linear transformation \citep{Duschinsky1937},
\begin{equation}
H_{m,\mathrm{full}}=\omega_{e}\sigma^{+}\sigma^{-}+\sum_{k}\left[\omega_{k}b_{k}^{\dagger}b_{k}+\lambda_{k}\sigma^{+}\sigma^{-}(b_{k}^{\dagger}+b_{k})\right],\label{eq:chain}
\end{equation}
yields the parameters $\{\omega_{k},\lambda_{k}\}$, or equivalently
the spectral density $J_{v}(\omega)=\sum_{k}\lambda_{k}^{2}\delta(\omega-\omega_{k})$,
determining the vibrational spectrum of the molecule. The single vibrational
mode in \autoref{eq:Ham} is then taken as the corresponding reaction
coordinate, with $\lambda_{v}=\sqrt{\sum_{k}\lambda_{k}^{2}}$ and
$\omega_{v}=\sum_{k}\omega_{k}\lambda_{k}^{2}/\lambda_{v}^{2}$~\citep{Chin2010,DelPino2018Dynamics}.

\begin{figure}
\includegraphics[width=1\columnwidth]{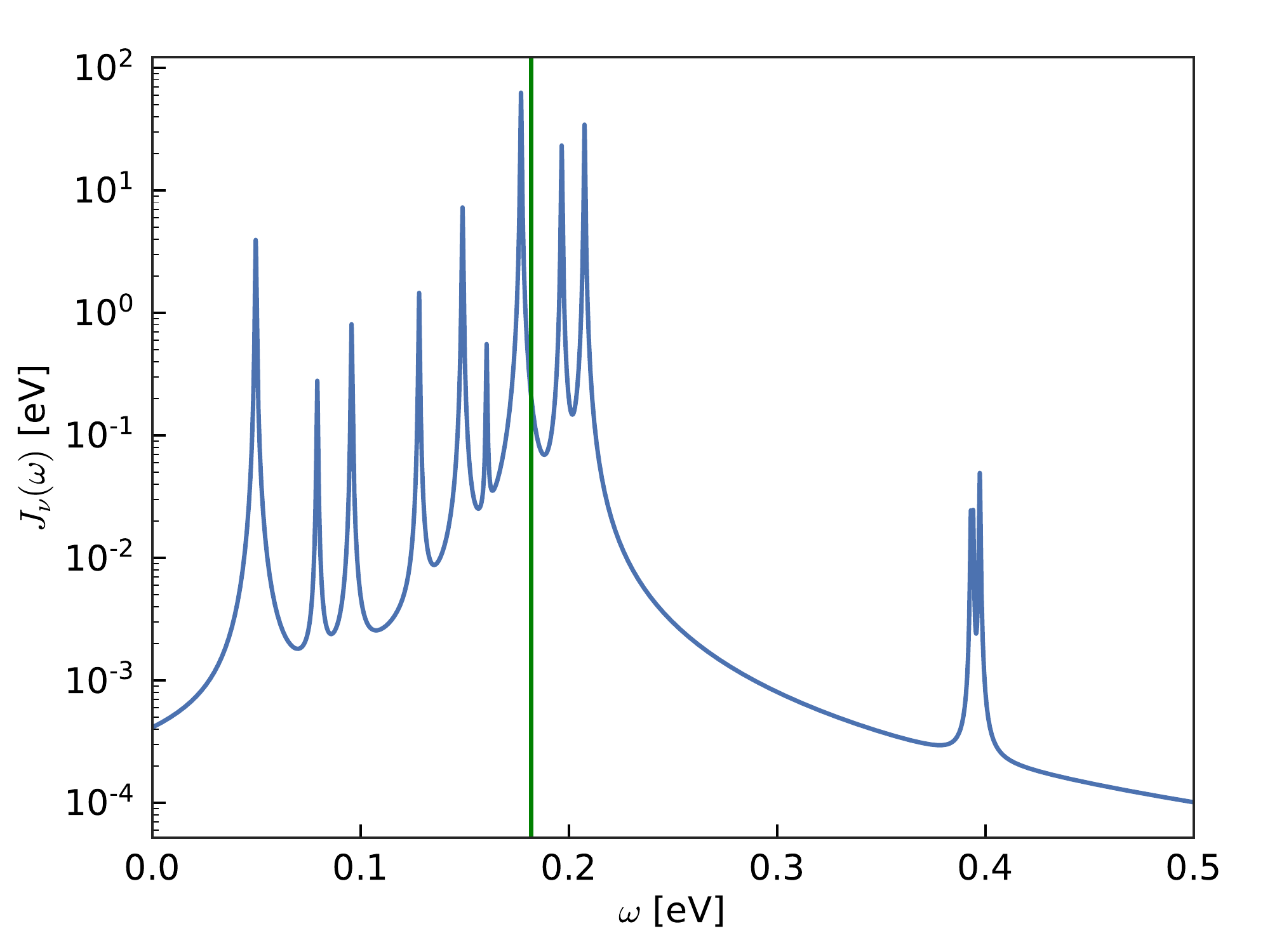} \caption{\label{fig:5}$J_{v}(\omega)$ for the anthracene molecule. The vertical
green line indicates the vibrational frequency $\omega_{v}$ of the
reaction coordinate.}
\end{figure}

In \autoref{fig:5}, we show the vibrational spectral density of
anthracene (convoluted with a Lorentzian to represent broadening due
to interactions with a solid-state environment). It can be seen that
$\omega_{v}$ is very close to the frequency of the dominant vibrational
mode in $J_{v}(\omega)$. We have additionally checked the validity
of the single-mode approximation by comparing the model calculations
above with time-dependent variational matrix product states (TDVMPS)
calculations~\citep{DelPino2018Dynamics,Schroder2019} in which the
full phononic spectral density, describing all vibrational modes of
the molecule and surroundings, is taken into account (see Supplementary
Information for details).

\section*{Data Availability}

The data that support the plots within this paper and other findings
of this study are available from the corresponding authors upon reasonable
request.

\section*{Acknowledgments}

We thank Alex~W.~Chin and Florian A.~Y.~N.~Schröder for their
help with the TDVMPS calculations, and Clàudia Climent for her help
with calculating the displaced harmonic oscillator model and with
the Duschinsky linear transformation. This work has been funded by
the European Research Council grant ERC-2016-STG-714870 and the Spanish
Ministry for Science, Innovation, and Universities -- AEI grants
RTI2018-099737-B-I00, PCI2018-093145 (through the QuantERA program
of the European Commission), and MDM-2014-0377 (through the María
de Maeztu program for Units of Excellence in R\&D).

\bibliography{references}

\begin{thebibliography}{59}%
\makeatletter
\providecommand \@ifxundefined [1]{%
 \@ifx{#1\undefined}
}%
\providecommand \@ifnum [1]{%
 \ifnum #1\expandafter \@firstoftwo
 \else \expandafter \@secondoftwo
 \fi
}%
\providecommand \@ifx [1]{%
 \ifx #1\expandafter \@firstoftwo
 \else \expandafter \@secondoftwo
 \fi
}%
\providecommand \natexlab [1]{#1}%
\providecommand \enquote  [1]{``#1''}%
\providecommand \bibnamefont  [1]{#1}%
\providecommand \bibfnamefont [1]{#1}%
\providecommand \citenamefont [1]{#1}%
\providecommand \href@noop [0]{\@secondoftwo}%
\providecommand \href [0]{\begingroup \@sanitize@url \@href}%
\providecommand \@href[1]{\@@startlink{#1}\@@href}%
\providecommand \@@href[1]{\endgroup#1\@@endlink}%
\providecommand \@sanitize@url [0]{\catcode `\\12\catcode `\$12\catcode
  `\&12\catcode `\#12\catcode `\^12\catcode `\_12\catcode `\%12\relax}%
\providecommand \@@startlink[1]{}%
\providecommand \@@endlink[0]{}%
\providecommand \url  [0]{\begingroup\@sanitize@url \@url }%
\providecommand \@url [1]{\endgroup\@href {#1}{\urlprefix }}%
\providecommand \urlprefix  [0]{URL }%
\providecommand \Eprint [0]{\href }%
\providecommand \doibase [0]{http://dx.doi.org/}%
\providecommand \selectlanguage [0]{\@gobble}%
\providecommand \bibinfo  [0]{\@secondoftwo}%
\providecommand \bibfield  [0]{\@secondoftwo}%
\providecommand \translation [1]{[#1]}%
\providecommand \BibitemOpen [0]{}%
\providecommand \bibitemStop [0]{}%
\providecommand \bibitemNoStop [0]{.\EOS\space}%
\providecommand \EOS [0]{\spacefactor3000\relax}%
\providecommand \BibitemShut  [1]{\csname bibitem#1\endcsname}%
\let\auto@bib@innerbib\@empty
\bibitem [{\citenamefont {Thompson}\ \emph {et~al.}(1992)\citenamefont
  {Thompson}, \citenamefont {Rempe},\ and\ \citenamefont
  {Kimble}}]{Thompson1992}%
  \BibitemOpen
  \bibfield  {author} {\bibinfo {author} {\bibfnamefont {R.~J.}\ \bibnamefont
  {Thompson}}, \bibinfo {author} {\bibfnamefont {G.}~\bibnamefont {Rempe}}, \
  and\ \bibinfo {author} {\bibfnamefont {H.~J.}\ \bibnamefont {Kimble}},\
  }\href {\doibase 10.1103/PhysRevLett.68.1132} {\bibfield  {journal} {\bibinfo
   {journal} {Phys. Rev. Lett.}\ }\textbf {\bibinfo {volume} {68}},\ \bibinfo
  {pages} {1132} (\bibinfo {year} {1992})}\BibitemShut {NoStop}%
\bibitem [{\citenamefont {Weisbuch}\ \emph {et~al.}(1992)\citenamefont
  {Weisbuch}, \citenamefont {Nishioka}, \citenamefont {Ishikawa},\ and\
  \citenamefont {Arakawa}}]{Weisbuch1992}%
  \BibitemOpen
  \bibfield  {author} {\bibinfo {author} {\bibfnamefont {C.}~\bibnamefont
  {Weisbuch}}, \bibinfo {author} {\bibfnamefont {M.}~\bibnamefont {Nishioka}},
  \bibinfo {author} {\bibfnamefont {A.}~\bibnamefont {Ishikawa}}, \ and\
  \bibinfo {author} {\bibfnamefont {Y.}~\bibnamefont {Arakawa}},\ }\href
  {\doibase 10.1103/PhysRevLett.69.3314} {\bibfield  {journal} {\bibinfo
  {journal} {Phys. Rev. Lett.}\ }\textbf {\bibinfo {volume} {69}},\ \bibinfo
  {pages} {3314} (\bibinfo {year} {1992})}\BibitemShut {NoStop}%
\bibitem [{\citenamefont {Lidzey}\ \emph {et~al.}(1998)\citenamefont {Lidzey},
  \citenamefont {Bradley}, \citenamefont {Skolnick}, \citenamefont {Virgili},
  \citenamefont {Walker},\ and\ \citenamefont {Whittaker}}]{Lidzey1998}%
  \BibitemOpen
  \bibfield  {author} {\bibinfo {author} {\bibfnamefont {D.~G.}\ \bibnamefont
  {Lidzey}}, \bibinfo {author} {\bibfnamefont {D.~D.~C.}\ \bibnamefont
  {Bradley}}, \bibinfo {author} {\bibfnamefont {M.~S.}\ \bibnamefont
  {Skolnick}}, \bibinfo {author} {\bibfnamefont {T.}~\bibnamefont {Virgili}},
  \bibinfo {author} {\bibfnamefont {S.}~\bibnamefont {Walker}}, \ and\ \bibinfo
  {author} {\bibfnamefont {D.~M.}\ \bibnamefont {Whittaker}},\ }\href {\doibase
  10.1038/25692} {\bibfield  {journal} {\bibinfo  {journal} {Nature}\ }\textbf
  {\bibinfo {volume} {395}},\ \bibinfo {pages} {53} (\bibinfo {year}
  {1998})}\BibitemShut {NoStop}%
\bibitem [{\citenamefont {Zengin}\ \emph {et~al.}(2015)\citenamefont {Zengin},
  \citenamefont {Wers{\"a}ll}, \citenamefont {Nilsson}, \citenamefont
  {Antosiewicz}, \citenamefont {K{\"a}ll},\ and\ \citenamefont
  {Shegai}}]{Zengin2015}%
  \BibitemOpen
  \bibfield  {author} {\bibinfo {author} {\bibfnamefont {G.}~\bibnamefont
  {Zengin}}, \bibinfo {author} {\bibfnamefont {M.}~\bibnamefont {Wers{\"a}ll}},
  \bibinfo {author} {\bibfnamefont {S.}~\bibnamefont {Nilsson}}, \bibinfo
  {author} {\bibfnamefont {T.~J.}\ \bibnamefont {Antosiewicz}}, \bibinfo
  {author} {\bibfnamefont {M.}~\bibnamefont {K{\"a}ll}}, \ and\ \bibinfo
  {author} {\bibfnamefont {T.}~\bibnamefont {Shegai}},\ }\href {\doibase
  10.1103/PhysRevLett.114.157401} {\bibfield  {journal} {\bibinfo  {journal}
  {Phys. Rev. Lett.}\ }\textbf {\bibinfo {volume} {114}},\ \bibinfo {pages}
  {157401} (\bibinfo {year} {2015})}\BibitemShut {NoStop}%
\bibitem [{\citenamefont {Chikkaraddy}\ \emph {et~al.}(2016)\citenamefont
  {Chikkaraddy}, \citenamefont {{de Nijs}}, \citenamefont {Benz}, \citenamefont
  {Barrow}, \citenamefont {Scherman}, \citenamefont {Rosta}, \citenamefont
  {Demetriadou}, \citenamefont {Fox}, \citenamefont {Hess},\ and\ \citenamefont
  {Baumberg}}]{Chikkaraddy2016}%
  \BibitemOpen
  \bibfield  {author} {\bibinfo {author} {\bibfnamefont {R.}~\bibnamefont
  {Chikkaraddy}}, \bibinfo {author} {\bibfnamefont {B.}~\bibnamefont {{de
  Nijs}}}, \bibinfo {author} {\bibfnamefont {F.}~\bibnamefont {Benz}}, \bibinfo
  {author} {\bibfnamefont {S.~J.}\ \bibnamefont {Barrow}}, \bibinfo {author}
  {\bibfnamefont {O.~A.}\ \bibnamefont {Scherman}}, \bibinfo {author}
  {\bibfnamefont {E.}~\bibnamefont {Rosta}}, \bibinfo {author} {\bibfnamefont
  {A.}~\bibnamefont {Demetriadou}}, \bibinfo {author} {\bibfnamefont
  {P.}~\bibnamefont {Fox}}, \bibinfo {author} {\bibfnamefont {O.}~\bibnamefont
  {Hess}}, \ and\ \bibinfo {author} {\bibfnamefont {J.~J.}\ \bibnamefont
  {Baumberg}},\ }\href {\doibase 10.1038/nature17974} {\bibfield  {journal}
  {\bibinfo  {journal} {Nature}\ }\textbf {\bibinfo {volume} {535}},\ \bibinfo
  {pages} {127} (\bibinfo {year} {2016})}\BibitemShut {NoStop}%
\bibitem [{\citenamefont {Ojambati}\ \emph {et~al.}(2019)\citenamefont
  {Ojambati}, \citenamefont {Chikkaraddy}, \citenamefont {Deacon},
  \citenamefont {Horton}, \citenamefont {Kos}, \citenamefont {Turek},
  \citenamefont {Keyser},\ and\ \citenamefont {Baumberg}}]{Ojambati2019}%
  \BibitemOpen
  \bibfield  {author} {\bibinfo {author} {\bibfnamefont {O.~S.}\ \bibnamefont
  {Ojambati}}, \bibinfo {author} {\bibfnamefont {R.}~\bibnamefont
  {Chikkaraddy}}, \bibinfo {author} {\bibfnamefont {W.~D.}\ \bibnamefont
  {Deacon}}, \bibinfo {author} {\bibfnamefont {M.}~\bibnamefont {Horton}},
  \bibinfo {author} {\bibfnamefont {D.}~\bibnamefont {Kos}}, \bibinfo {author}
  {\bibfnamefont {V.~A.}\ \bibnamefont {Turek}}, \bibinfo {author}
  {\bibfnamefont {U.~F.}\ \bibnamefont {Keyser}}, \ and\ \bibinfo {author}
  {\bibfnamefont {J.~J.}\ \bibnamefont {Baumberg}},\ }\href {\doibase
  10.1038/s41467-019-08611-5} {\bibfield  {journal} {\bibinfo  {journal} {Nat.
  Commun.}\ }\textbf {\bibinfo {volume} {10}},\ \bibinfo {pages} {1049}
  (\bibinfo {year} {2019})}\BibitemShut {NoStop}%
\bibitem [{\citenamefont {Vasa}\ \emph {et~al.}(2013)\citenamefont {Vasa},
  \citenamefont {Wang}, \citenamefont {Pomraenke}, \citenamefont {Lammers},
  \citenamefont {Maiuri}, \citenamefont {Manzoni}, \citenamefont {Cerullo},\
  and\ \citenamefont {Lienau}}]{Vasa2013}%
  \BibitemOpen
  \bibfield  {author} {\bibinfo {author} {\bibfnamefont {P.}~\bibnamefont
  {Vasa}}, \bibinfo {author} {\bibfnamefont {W.}~\bibnamefont {Wang}}, \bibinfo
  {author} {\bibfnamefont {R.}~\bibnamefont {Pomraenke}}, \bibinfo {author}
  {\bibfnamefont {M.}~\bibnamefont {Lammers}}, \bibinfo {author} {\bibfnamefont
  {M.}~\bibnamefont {Maiuri}}, \bibinfo {author} {\bibfnamefont
  {C.}~\bibnamefont {Manzoni}}, \bibinfo {author} {\bibfnamefont
  {G.}~\bibnamefont {Cerullo}}, \ and\ \bibinfo {author} {\bibfnamefont
  {C.}~\bibnamefont {Lienau}},\ }\href {\doibase 10.1038/nphoton.2012.340}
  {\bibfield  {journal} {\bibinfo  {journal} {Nat. Photonics}\ }\textbf
  {\bibinfo {volume} {7}},\ \bibinfo {pages} {128} (\bibinfo {year}
  {2013})}\BibitemShut {NoStop}%
\bibitem [{\citenamefont {T{\"o}rm{\"a}}\ and\ \citenamefont
  {Barnes}(2015)}]{Torma2015}%
  \BibitemOpen
  \bibfield  {author} {\bibinfo {author} {\bibfnamefont {P.}~\bibnamefont
  {T{\"o}rm{\"a}}}\ and\ \bibinfo {author} {\bibfnamefont {W.~L.}\ \bibnamefont
  {Barnes}},\ }\href {\doibase 10.1088/0034-4885/78/1/013901} {\bibfield
  {journal} {\bibinfo  {journal} {Rep. Prog. Phys.}\ }\textbf {\bibinfo
  {volume} {78}},\ \bibinfo {pages} {013901} (\bibinfo {year}
  {2015})}\BibitemShut {NoStop}%
\bibitem [{\citenamefont {{\'C}wik}\ \emph {et~al.}(2016)\citenamefont
  {{\'C}wik}, \citenamefont {Kirton}, \citenamefont {De~Liberato},\ and\
  \citenamefont {Keeling}}]{Cwik2016}%
  \BibitemOpen
  \bibfield  {author} {\bibinfo {author} {\bibfnamefont {J.~A.}\ \bibnamefont
  {{\'C}wik}}, \bibinfo {author} {\bibfnamefont {P.}~\bibnamefont {Kirton}},
  \bibinfo {author} {\bibfnamefont {S.}~\bibnamefont {De~Liberato}}, \ and\
  \bibinfo {author} {\bibfnamefont {J.}~\bibnamefont {Keeling}},\ }\href
  {\doibase 10.1103/PhysRevA.93.033840} {\bibfield  {journal} {\bibinfo
  {journal} {Phys. Rev. A}\ }\textbf {\bibinfo {volume} {93}},\ \bibinfo
  {pages} {033840} (\bibinfo {year} {2016})}\BibitemShut {NoStop}%
\bibitem [{\citenamefont {{del Pino}}\ \emph {et~al.}(2018)\citenamefont {{del
  Pino}}, \citenamefont {Schr{\"o}der}, \citenamefont {Chin}, \citenamefont
  {Feist},\ and\ \citenamefont {{Garcia-Vidal}}}]{DelPino2018Dynamics}%
  \BibitemOpen
  \bibfield  {author} {\bibinfo {author} {\bibfnamefont {J.}~\bibnamefont {{del
  Pino}}}, \bibinfo {author} {\bibfnamefont {F.~A. Y.~N.}\ \bibnamefont
  {Schr{\"o}der}}, \bibinfo {author} {\bibfnamefont {A.~W.}\ \bibnamefont
  {Chin}}, \bibinfo {author} {\bibfnamefont {J.}~\bibnamefont {Feist}}, \ and\
  \bibinfo {author} {\bibfnamefont {F.~J.}\ \bibnamefont {{Garcia-Vidal}}},\
  }\href {\doibase 10.1103/PhysRevLett.121.227401} {\bibfield  {journal}
  {\bibinfo  {journal} {Phys. Rev. Lett.}\ }\textbf {\bibinfo {volume} {121}},\
  \bibinfo {pages} {227401} (\bibinfo {year} {2018})}\BibitemShut {NoStop}%
\bibitem [{\citenamefont {Herrera}\ and\ \citenamefont
  {Spano}(2018)}]{Herrera2018Theory}%
  \BibitemOpen
  \bibfield  {author} {\bibinfo {author} {\bibfnamefont {F.}~\bibnamefont
  {Herrera}}\ and\ \bibinfo {author} {\bibfnamefont {F.~C.}\ \bibnamefont
  {Spano}},\ }\href {\doibase 10.1021/acsphotonics.7b00728} {\bibfield
  {journal} {\bibinfo  {journal} {ACS Photonics}\ }\textbf {\bibinfo {volume}
  {5}},\ \bibinfo {pages} {65} (\bibinfo {year} {2018})}\BibitemShut {NoStop}%
\bibitem [{\citenamefont {Singh}\ \emph {et~al.}(2018)\citenamefont {Singh},
  \citenamefont {{de Roque}}, \citenamefont {Calbris}, \citenamefont {Hugall},\
  and\ \citenamefont {{van Hulst}}}]{Singh2018}%
  \BibitemOpen
  \bibfield  {author} {\bibinfo {author} {\bibfnamefont {A.}~\bibnamefont
  {Singh}}, \bibinfo {author} {\bibfnamefont {P.~M.}\ \bibnamefont {{de
  Roque}}}, \bibinfo {author} {\bibfnamefont {G.}~\bibnamefont {Calbris}},
  \bibinfo {author} {\bibfnamefont {J.~T.}\ \bibnamefont {Hugall}}, \ and\
  \bibinfo {author} {\bibfnamefont {N.~F.}\ \bibnamefont {{van Hulst}}},\
  }\href {\doibase 10.1021/acs.nanolett.8b00239} {\bibfield  {journal}
  {\bibinfo  {journal} {Nano Lett.}\ }\textbf {\bibinfo {volume} {18}},\
  \bibinfo {pages} {2538} (\bibinfo {year} {2018})}\BibitemShut {NoStop}%
\bibitem [{\citenamefont {Coles}\ \emph {et~al.}(2014)\citenamefont {Coles},
  \citenamefont {Somaschi}, \citenamefont {Michetti}, \citenamefont {Clark},
  \citenamefont {Lagoudakis}, \citenamefont {Savvidis},\ and\ \citenamefont
  {Lidzey}}]{Coles2014}%
  \BibitemOpen
  \bibfield  {author} {\bibinfo {author} {\bibfnamefont {D.~M.}\ \bibnamefont
  {Coles}}, \bibinfo {author} {\bibfnamefont {N.}~\bibnamefont {Somaschi}},
  \bibinfo {author} {\bibfnamefont {P.}~\bibnamefont {Michetti}}, \bibinfo
  {author} {\bibfnamefont {C.}~\bibnamefont {Clark}}, \bibinfo {author}
  {\bibfnamefont {P.~G.}\ \bibnamefont {Lagoudakis}}, \bibinfo {author}
  {\bibfnamefont {P.~G.}\ \bibnamefont {Savvidis}}, \ and\ \bibinfo {author}
  {\bibfnamefont {D.~G.}\ \bibnamefont {Lidzey}},\ }\href {\doibase
  10.1038/nmat3950} {\bibfield  {journal} {\bibinfo  {journal} {Nat. Mater.}\
  }\textbf {\bibinfo {volume} {13}},\ \bibinfo {pages} {712} (\bibinfo {year}
  {2014})}\BibitemShut {NoStop}%
\bibitem [{\citenamefont {Orgiu}\ \emph {et~al.}(2015)\citenamefont {Orgiu},
  \citenamefont {George}, \citenamefont {Hutchison}, \citenamefont {Devaux},
  \citenamefont {Dayen}, \citenamefont {Doudin}, \citenamefont {Stellacci},
  \citenamefont {Genet}, \citenamefont {Schachenmayer}, \citenamefont {Genes},
  \citenamefont {Pupillo}, \citenamefont {Samor{\`i}},\ and\ \citenamefont
  {Ebbesen}}]{Orgiu2015}%
  \BibitemOpen
  \bibfield  {author} {\bibinfo {author} {\bibfnamefont {E.}~\bibnamefont
  {Orgiu}}, \bibinfo {author} {\bibfnamefont {J.}~\bibnamefont {George}},
  \bibinfo {author} {\bibfnamefont {J.~A.}\ \bibnamefont {Hutchison}}, \bibinfo
  {author} {\bibfnamefont {E.}~\bibnamefont {Devaux}}, \bibinfo {author}
  {\bibfnamefont {J.~F.}\ \bibnamefont {Dayen}}, \bibinfo {author}
  {\bibfnamefont {B.}~\bibnamefont {Doudin}}, \bibinfo {author} {\bibfnamefont
  {F.}~\bibnamefont {Stellacci}}, \bibinfo {author} {\bibfnamefont
  {C.}~\bibnamefont {Genet}}, \bibinfo {author} {\bibfnamefont
  {J.}~\bibnamefont {Schachenmayer}}, \bibinfo {author} {\bibfnamefont
  {C.}~\bibnamefont {Genes}}, \bibinfo {author} {\bibfnamefont
  {G.}~\bibnamefont {Pupillo}}, \bibinfo {author} {\bibfnamefont
  {P.}~\bibnamefont {Samor{\`i}}}, \ and\ \bibinfo {author} {\bibfnamefont
  {T.~W.}\ \bibnamefont {Ebbesen}},\ }\href {\doibase 10.1038/nmat4392}
  {\bibfield  {journal} {\bibinfo  {journal} {Nat. Mater.}\ }\textbf {\bibinfo
  {volume} {14}},\ \bibinfo {pages} {1123} (\bibinfo {year}
  {2015})}\BibitemShut {NoStop}%
\bibitem [{\citenamefont {Feist}\ and\ \citenamefont
  {{Garcia-Vidal}}(2015)}]{Feist2015}%
  \BibitemOpen
  \bibfield  {author} {\bibinfo {author} {\bibfnamefont {J.}~\bibnamefont
  {Feist}}\ and\ \bibinfo {author} {\bibfnamefont {F.~J.}\ \bibnamefont
  {{Garcia-Vidal}}},\ }\href {\doibase 10.1103/PhysRevLett.114.196402}
  {\bibfield  {journal} {\bibinfo  {journal} {Phys. Rev. Lett.}\ }\textbf
  {\bibinfo {volume} {114}},\ \bibinfo {pages} {196402} (\bibinfo {year}
  {2015})}\BibitemShut {NoStop}%
\bibitem [{\citenamefont {Zhong}\ \emph {et~al.}(2017)\citenamefont {Zhong},
  \citenamefont {Chervy}, \citenamefont {Zhang}, \citenamefont {Thomas},
  \citenamefont {George}, \citenamefont {Genet}, \citenamefont {Hutchison},\
  and\ \citenamefont {Ebbesen}}]{Zhong2017}%
  \BibitemOpen
  \bibfield  {author} {\bibinfo {author} {\bibfnamefont {X.}~\bibnamefont
  {Zhong}}, \bibinfo {author} {\bibfnamefont {T.}~\bibnamefont {Chervy}},
  \bibinfo {author} {\bibfnamefont {L.}~\bibnamefont {Zhang}}, \bibinfo
  {author} {\bibfnamefont {A.}~\bibnamefont {Thomas}}, \bibinfo {author}
  {\bibfnamefont {J.}~\bibnamefont {George}}, \bibinfo {author} {\bibfnamefont
  {C.}~\bibnamefont {Genet}}, \bibinfo {author} {\bibfnamefont {J.~A.}\
  \bibnamefont {Hutchison}}, \ and\ \bibinfo {author} {\bibfnamefont {T.~W.}\
  \bibnamefont {Ebbesen}},\ }\href {\doibase 10.1002/anie.201703539} {\bibfield
   {journal} {\bibinfo  {journal} {Angew. Chem. Int. Ed.}\ }\textbf {\bibinfo
  {volume} {56}},\ \bibinfo {pages} {9034} (\bibinfo {year}
  {2017})}\BibitemShut {NoStop}%
\bibitem [{\citenamefont {Hutchison}\ \emph {et~al.}(2012)\citenamefont
  {Hutchison}, \citenamefont {Schwartz}, \citenamefont {Genet}, \citenamefont
  {Devaux},\ and\ \citenamefont {Ebbesen}}]{Hutchison2012}%
  \BibitemOpen
  \bibfield  {author} {\bibinfo {author} {\bibfnamefont {J.~A.}\ \bibnamefont
  {Hutchison}}, \bibinfo {author} {\bibfnamefont {T.}~\bibnamefont {Schwartz}},
  \bibinfo {author} {\bibfnamefont {C.}~\bibnamefont {Genet}}, \bibinfo
  {author} {\bibfnamefont {E.}~\bibnamefont {Devaux}}, \ and\ \bibinfo {author}
  {\bibfnamefont {T.~W.}\ \bibnamefont {Ebbesen}},\ }\href {\doibase
  10.1002/ange.201107033} {\bibfield  {journal} {\bibinfo  {journal} {Angew.
  Chem.}\ }\textbf {\bibinfo {volume} {124}},\ \bibinfo {pages} {1624}
  (\bibinfo {year} {2012})}\BibitemShut {NoStop}%
\bibitem [{\citenamefont {Galego}\ \emph {et~al.}(2015)\citenamefont {Galego},
  \citenamefont {{Garcia-Vidal}},\ and\ \citenamefont {Feist}}]{Galego2015}%
  \BibitemOpen
  \bibfield  {author} {\bibinfo {author} {\bibfnamefont {J.}~\bibnamefont
  {Galego}}, \bibinfo {author} {\bibfnamefont {F.~J.}\ \bibnamefont
  {{Garcia-Vidal}}}, \ and\ \bibinfo {author} {\bibfnamefont {J.}~\bibnamefont
  {Feist}},\ }\href {\doibase 10.1103/PhysRevX.5.041022} {\bibfield  {journal}
  {\bibinfo  {journal} {Phys. Rev. X}\ }\textbf {\bibinfo {volume} {5}},\
  \bibinfo {pages} {041022} (\bibinfo {year} {2015})}\BibitemShut {NoStop}%
\bibitem [{\citenamefont {Herrera}\ and\ \citenamefont
  {Spano}(2016)}]{Herrera2016}%
  \BibitemOpen
  \bibfield  {author} {\bibinfo {author} {\bibfnamefont {F.}~\bibnamefont
  {Herrera}}\ and\ \bibinfo {author} {\bibfnamefont {F.~C.}\ \bibnamefont
  {Spano}},\ }\href {\doibase 10.1103/PhysRevLett.116.238301} {\bibfield
  {journal} {\bibinfo  {journal} {Phys. Rev. Lett.}\ }\textbf {\bibinfo
  {volume} {116}},\ \bibinfo {pages} {238301} (\bibinfo {year}
  {2016})}\BibitemShut {NoStop}%
\bibitem [{\citenamefont {Thomas}\ \emph {et~al.}(2016)\citenamefont {Thomas},
  \citenamefont {George}, \citenamefont {Shalabney}, \citenamefont {Dryzhakov},
  \citenamefont {Varma}, \citenamefont {Moran}, \citenamefont {Chervy},
  \citenamefont {Zhong}, \citenamefont {Devaux}, \citenamefont {Genet},
  \citenamefont {Hutchison},\ and\ \citenamefont {Ebbesen}}]{Thomas2016}%
  \BibitemOpen
  \bibfield  {author} {\bibinfo {author} {\bibfnamefont {A.}~\bibnamefont
  {Thomas}}, \bibinfo {author} {\bibfnamefont {J.}~\bibnamefont {George}},
  \bibinfo {author} {\bibfnamefont {A.}~\bibnamefont {Shalabney}}, \bibinfo
  {author} {\bibfnamefont {M.}~\bibnamefont {Dryzhakov}}, \bibinfo {author}
  {\bibfnamefont {S.~J.}\ \bibnamefont {Varma}}, \bibinfo {author}
  {\bibfnamefont {J.}~\bibnamefont {Moran}}, \bibinfo {author} {\bibfnamefont
  {T.}~\bibnamefont {Chervy}}, \bibinfo {author} {\bibfnamefont
  {X.}~\bibnamefont {Zhong}}, \bibinfo {author} {\bibfnamefont
  {E.}~\bibnamefont {Devaux}}, \bibinfo {author} {\bibfnamefont
  {C.}~\bibnamefont {Genet}}, \bibinfo {author} {\bibfnamefont {J.~A.}\
  \bibnamefont {Hutchison}}, \ and\ \bibinfo {author} {\bibfnamefont {T.~W.}\
  \bibnamefont {Ebbesen}},\ }\href {\doibase 10.1002/anie.201605504} {\bibfield
   {journal} {\bibinfo  {journal} {Angew. Chem. Int. Ed.}\ }\textbf {\bibinfo
  {volume} {55}},\ \bibinfo {pages} {11462} (\bibinfo {year}
  {2016})}\BibitemShut {NoStop}%
\bibitem [{\citenamefont {Flick}\ \emph {et~al.}(2017)\citenamefont {Flick},
  \citenamefont {Ruggenthaler}, \citenamefont {Appel},\ and\ \citenamefont
  {Rubio}}]{Flick2017Atoms}%
  \BibitemOpen
  \bibfield  {author} {\bibinfo {author} {\bibfnamefont {J.}~\bibnamefont
  {Flick}}, \bibinfo {author} {\bibfnamefont {M.}~\bibnamefont {Ruggenthaler}},
  \bibinfo {author} {\bibfnamefont {H.}~\bibnamefont {Appel}}, \ and\ \bibinfo
  {author} {\bibfnamefont {A.}~\bibnamefont {Rubio}},\ }\href {\doibase
  10.1073/pnas.1615509114} {\bibfield  {journal} {\bibinfo  {journal} {Proc.
  Natl. Acad. Sci.}\ }\textbf {\bibinfo {volume} {114}},\ \bibinfo {pages}
  {3026} (\bibinfo {year} {2017})}\BibitemShut {NoStop}%
\bibitem [{\citenamefont {Munkhbat}\ \emph {et~al.}(2018)\citenamefont
  {Munkhbat}, \citenamefont {Wers{\"a}ll}, \citenamefont {Baranov},
  \citenamefont {Antosiewicz},\ and\ \citenamefont {Shegai}}]{Munkhbat2018}%
  \BibitemOpen
  \bibfield  {author} {\bibinfo {author} {\bibfnamefont {B.}~\bibnamefont
  {Munkhbat}}, \bibinfo {author} {\bibfnamefont {M.}~\bibnamefont
  {Wers{\"a}ll}}, \bibinfo {author} {\bibfnamefont {D.~G.}\ \bibnamefont
  {Baranov}}, \bibinfo {author} {\bibfnamefont {T.~J.}\ \bibnamefont
  {Antosiewicz}}, \ and\ \bibinfo {author} {\bibfnamefont {T.}~\bibnamefont
  {Shegai}},\ }\href {\doibase 10.1126/sciadv.aas9552} {\bibfield  {journal}
  {\bibinfo  {journal} {Sci. Adv.}\ }\textbf {\bibinfo {volume} {4}},\ \bibinfo
  {pages} {eaas9552} (\bibinfo {year} {2018})}\BibitemShut {NoStop}%
\bibitem [{\citenamefont {Fregoni}\ \emph {et~al.}(2018)\citenamefont
  {Fregoni}, \citenamefont {Granucci}, \citenamefont {Coccia}, \citenamefont
  {Persico},\ and\ \citenamefont {Corni}}]{Fregoni2018}%
  \BibitemOpen
  \bibfield  {author} {\bibinfo {author} {\bibfnamefont {J.}~\bibnamefont
  {Fregoni}}, \bibinfo {author} {\bibfnamefont {G.}~\bibnamefont {Granucci}},
  \bibinfo {author} {\bibfnamefont {E.}~\bibnamefont {Coccia}}, \bibinfo
  {author} {\bibfnamefont {M.}~\bibnamefont {Persico}}, \ and\ \bibinfo
  {author} {\bibfnamefont {S.}~\bibnamefont {Corni}},\ }\href {\doibase
  10.1038/s41467-018-06971-y} {\bibfield  {journal} {\bibinfo  {journal} {Nat.
  Commun.}\ }\textbf {\bibinfo {volume} {9}},\ \bibinfo {pages} {4688}
  (\bibinfo {year} {2018})}\BibitemShut {NoStop}%
\bibitem [{\citenamefont {Peters}\ \emph {et~al.}(2019)\citenamefont {Peters},
  \citenamefont {Faruk}, \citenamefont {Asane}, \citenamefont {Alexander},
  \citenamefont {Peters}, \citenamefont {Prayakarao}, \citenamefont {Rout},\
  and\ \citenamefont {Noginov}}]{Peters2019}%
  \BibitemOpen
  \bibfield  {author} {\bibinfo {author} {\bibfnamefont {V.~N.}\ \bibnamefont
  {Peters}}, \bibinfo {author} {\bibfnamefont {M.~O.}\ \bibnamefont {Faruk}},
  \bibinfo {author} {\bibfnamefont {J.}~\bibnamefont {Asane}}, \bibinfo
  {author} {\bibfnamefont {R.}~\bibnamefont {Alexander}}, \bibinfo {author}
  {\bibfnamefont {D.~A.}\ \bibnamefont {Peters}}, \bibinfo {author}
  {\bibfnamefont {S.}~\bibnamefont {Prayakarao}}, \bibinfo {author}
  {\bibfnamefont {S.}~\bibnamefont {Rout}}, \ and\ \bibinfo {author}
  {\bibfnamefont {M.~A.}\ \bibnamefont {Noginov}},\ }\href {\doibase
  10.1364/OPTICA.6.000318} {\bibfield  {journal} {\bibinfo  {journal} {Optica}\
  }\textbf {\bibinfo {volume} {6}},\ \bibinfo {pages} {318} (\bibinfo {year}
  {2019})}\BibitemShut {NoStop}%
\bibitem [{\citenamefont {Du}\ \emph {et~al.}(2019)\citenamefont {Du},
  \citenamefont {Ribeiro},\ and\ \citenamefont {{Yuen-Zhou}}}]{Du2019Remote}%
  \BibitemOpen
  \bibfield  {author} {\bibinfo {author} {\bibfnamefont {M.}~\bibnamefont
  {Du}}, \bibinfo {author} {\bibfnamefont {R.~F.}\ \bibnamefont {Ribeiro}}, \
  and\ \bibinfo {author} {\bibfnamefont {J.}~\bibnamefont {{Yuen-Zhou}}},\
  }\href {\doibase 10.1016/j.chempr.2019.02.009} {\bibfield  {journal}
  {\bibinfo  {journal} {Chem}\ }\textbf {\bibinfo {volume} {5}},\ \bibinfo
  {pages} {1167} (\bibinfo {year} {2019})}\BibitemShut {NoStop}%
\bibitem [{\citenamefont {Stranius}\ \emph {et~al.}(2018)\citenamefont
  {Stranius}, \citenamefont {Hertzog},\ and\ \citenamefont
  {B{\"o}rjesson}}]{Stranius2018}%
  \BibitemOpen
  \bibfield  {author} {\bibinfo {author} {\bibfnamefont {K.}~\bibnamefont
  {Stranius}}, \bibinfo {author} {\bibfnamefont {M.}~\bibnamefont {Hertzog}}, \
  and\ \bibinfo {author} {\bibfnamefont {K.}~\bibnamefont {B{\"o}rjesson}},\
  }\href {\doibase 10.1038/s41467-018-04736-1} {\bibfield  {journal} {\bibinfo
  {journal} {Nat. Commun.}\ }\textbf {\bibinfo {volume} {9}},\ \bibinfo {pages}
  {2273} (\bibinfo {year} {2018})}\BibitemShut {NoStop}%
\bibitem [{\citenamefont {{Mart{\'i}nez-Mart{\'i}nez}}\ \emph
  {et~al.}(2018)\citenamefont {{Mart{\'i}nez-Mart{\'i}nez}}, \citenamefont
  {Du}, \citenamefont {F.~Ribeiro}, \citenamefont {{K{\'e}na-Cohen}},\ and\
  \citenamefont {{Yuen-Zhou}}}]{Martinez-Martinez2018Polariton}%
  \BibitemOpen
  \bibfield  {author} {\bibinfo {author} {\bibfnamefont {L.~A.}\ \bibnamefont
  {{Mart{\'i}nez-Mart{\'i}nez}}}, \bibinfo {author} {\bibfnamefont
  {M.}~\bibnamefont {Du}}, \bibinfo {author} {\bibfnamefont {R.}~\bibnamefont
  {F.~Ribeiro}}, \bibinfo {author} {\bibfnamefont {S.}~\bibnamefont
  {{K{\'e}na-Cohen}}}, \ and\ \bibinfo {author} {\bibfnamefont
  {J.}~\bibnamefont {{Yuen-Zhou}}},\ }\href {\doibase
  10.1021/acs.jpclett.8b00008} {\bibfield  {journal} {\bibinfo  {journal} {J.
  Phys. Chem. Lett.}\ }\textbf {\bibinfo {volume} {9}},\ \bibinfo {pages}
  {1951} (\bibinfo {year} {2018})}\BibitemShut {NoStop}%
\bibitem [{\citenamefont {Zewail}(2000)}]{Zewail2000}%
  \BibitemOpen
  \bibfield  {author} {\bibinfo {author} {\bibfnamefont {A.~H.}\ \bibnamefont
  {Zewail}},\ }\href {\doibase 10.1021/jp001460h} {\bibfield  {journal}
  {\bibinfo  {journal} {J. Phys. Chem. A}\ }\textbf {\bibinfo {volume} {104}},\
  \bibinfo {pages} {5660} (\bibinfo {year} {2000})}\BibitemShut {NoStop}%
\bibitem [{\citenamefont {Krausz}\ and\ \citenamefont
  {Ivanov}(2009)}]{Krausz2009}%
  \BibitemOpen
  \bibfield  {author} {\bibinfo {author} {\bibfnamefont {F.}~\bibnamefont
  {Krausz}}\ and\ \bibinfo {author} {\bibfnamefont {M.}~\bibnamefont
  {Ivanov}},\ }\href {\doibase 10.1103/RevModPhys.81.163} {\bibfield  {journal}
  {\bibinfo  {journal} {Rev. Mod. Phys.}\ }\textbf {\bibinfo {volume} {81}},\
  \bibinfo {pages} {163} (\bibinfo {year} {2009})}\BibitemShut {NoStop}%
\bibitem [{\citenamefont {Potter}\ \emph {et~al.}(1992)\citenamefont {Potter},
  \citenamefont {Herek}, \citenamefont {Pedersen}, \citenamefont {Liu},\ and\
  \citenamefont {Zewail}}]{Potter1992}%
  \BibitemOpen
  \bibfield  {author} {\bibinfo {author} {\bibfnamefont {E.~D.}\ \bibnamefont
  {Potter}}, \bibinfo {author} {\bibfnamefont {J.~L.}\ \bibnamefont {Herek}},
  \bibinfo {author} {\bibfnamefont {S.}~\bibnamefont {Pedersen}}, \bibinfo
  {author} {\bibfnamefont {Q.}~\bibnamefont {Liu}}, \ and\ \bibinfo {author}
  {\bibfnamefont {A.~H.}\ \bibnamefont {Zewail}},\ }\href {\doibase
  10.1038/355066a0} {\bibfield  {journal} {\bibinfo  {journal} {Nature}\
  }\textbf {\bibinfo {volume} {355}},\ \bibinfo {pages} {66} (\bibinfo {year}
  {1992})}\BibitemShut {NoStop}%
\bibitem [{\citenamefont {Assion}\ \emph {et~al.}(1998)\citenamefont {Assion},
  \citenamefont {Baumert}, \citenamefont {Bergt}, \citenamefont {Brixner},
  \citenamefont {Kiefer}, \citenamefont {Seyfried}, \citenamefont {Strehle},\
  and\ \citenamefont {Gerber}}]{Assion1998}%
  \BibitemOpen
  \bibfield  {author} {\bibinfo {author} {\bibfnamefont {A.}~\bibnamefont
  {Assion}}, \bibinfo {author} {\bibfnamefont {T.}~\bibnamefont {Baumert}},
  \bibinfo {author} {\bibfnamefont {M.}~\bibnamefont {Bergt}}, \bibinfo
  {author} {\bibfnamefont {T.}~\bibnamefont {Brixner}}, \bibinfo {author}
  {\bibfnamefont {B.}~\bibnamefont {Kiefer}}, \bibinfo {author} {\bibfnamefont
  {V.}~\bibnamefont {Seyfried}}, \bibinfo {author} {\bibfnamefont
  {M.}~\bibnamefont {Strehle}}, \ and\ \bibinfo {author} {\bibfnamefont
  {G.}~\bibnamefont {Gerber}},\ }\href {\doibase 10.1126/science.282.5390.919}
  {\bibfield  {journal} {\bibinfo  {journal} {Science}\ }\textbf {\bibinfo
  {volume} {282}},\ \bibinfo {pages} {919} (\bibinfo {year}
  {1998})}\BibitemShut {NoStop}%
\bibitem [{\citenamefont {Kling}\ \emph {et~al.}(2006)\citenamefont {Kling},
  \citenamefont {Siedschlag}, \citenamefont {Verhoef}, \citenamefont {Khan},
  \citenamefont {Schultze}, \citenamefont {Uphues}, \citenamefont {Ni},
  \citenamefont {Uiberacker}, \citenamefont {Drescher}, \citenamefont
  {Krausz},\ and\ \citenamefont {Vrakking}}]{Kling2006}%
  \BibitemOpen
  \bibfield  {author} {\bibinfo {author} {\bibfnamefont {M.~F.}\ \bibnamefont
  {Kling}}, \bibinfo {author} {\bibfnamefont {C.}~\bibnamefont {Siedschlag}},
  \bibinfo {author} {\bibfnamefont {A.~J.}\ \bibnamefont {Verhoef}}, \bibinfo
  {author} {\bibfnamefont {J.~I.}\ \bibnamefont {Khan}}, \bibinfo {author}
  {\bibfnamefont {M.}~\bibnamefont {Schultze}}, \bibinfo {author}
  {\bibfnamefont {T.}~\bibnamefont {Uphues}}, \bibinfo {author} {\bibfnamefont
  {Y.}~\bibnamefont {Ni}}, \bibinfo {author} {\bibfnamefont {M.}~\bibnamefont
  {Uiberacker}}, \bibinfo {author} {\bibfnamefont {M.}~\bibnamefont
  {Drescher}}, \bibinfo {author} {\bibfnamefont {F.}~\bibnamefont {Krausz}}, \
  and\ \bibinfo {author} {\bibfnamefont {M.~J.~J.}\ \bibnamefont {Vrakking}},\
  }\href {\doibase doi: 10.1126/science.1126259} {\bibfield  {journal}
  {\bibinfo  {journal} {Science}\ }\textbf {\bibinfo {volume} {312}},\ \bibinfo
  {pages} {246} (\bibinfo {year} {2006})}\BibitemShut {NoStop}%
\bibitem [{\citenamefont {Corrales}\ \emph {et~al.}(2014)\citenamefont
  {Corrales}, \citenamefont {{Gonz{\'a}lez-V{\'a}zquez}}, \citenamefont
  {Balerdi}, \citenamefont {Sol{\'a}}, \citenamefont {{de Nalda}},\ and\
  \citenamefont {Ba{\~n}ares}}]{Corrales2014}%
  \BibitemOpen
  \bibfield  {author} {\bibinfo {author} {\bibfnamefont {M.~E.}\ \bibnamefont
  {Corrales}}, \bibinfo {author} {\bibfnamefont {J.}~\bibnamefont
  {{Gonz{\'a}lez-V{\'a}zquez}}}, \bibinfo {author} {\bibfnamefont
  {G.}~\bibnamefont {Balerdi}}, \bibinfo {author} {\bibfnamefont {I.~R.}\
  \bibnamefont {Sol{\'a}}}, \bibinfo {author} {\bibfnamefont {R.}~\bibnamefont
  {{de Nalda}}}, \ and\ \bibinfo {author} {\bibfnamefont {L.}~\bibnamefont
  {Ba{\~n}ares}},\ }\href {\doibase 10.1038/nchem.2006} {\bibfield  {journal}
  {\bibinfo  {journal} {Nat. Chem.}\ }\textbf {\bibinfo {volume} {6}},\
  \bibinfo {pages} {785} (\bibinfo {year} {2014})}\BibitemShut {NoStop}%
\bibitem [{\citenamefont {Palacios}\ and\ \citenamefont
  {Mart{\'i}n}(2019)}]{Palacios2019}%
  \BibitemOpen
  \bibfield  {author} {\bibinfo {author} {\bibfnamefont {A.}~\bibnamefont
  {Palacios}}\ and\ \bibinfo {author} {\bibfnamefont {F.}~\bibnamefont
  {Mart{\'i}n}},\ }\href {\doibase 10.1002/wcms.1430} {\bibfield  {journal}
  {\bibinfo  {journal} {WIREs: Computational Molecular Science}\ }\textbf
  {\bibinfo {volume} {10}},\ \bibinfo {pages} {e1430} (\bibinfo {year}
  {2019})}\BibitemShut {NoStop}%
\bibitem [{\citenamefont {Kowalewski}\ \emph {et~al.}(2016)\citenamefont
  {Kowalewski}, \citenamefont {Bennett},\ and\ \citenamefont
  {Mukamel}}]{Kowalewski2016Cavity}%
  \BibitemOpen
  \bibfield  {author} {\bibinfo {author} {\bibfnamefont {M.}~\bibnamefont
  {Kowalewski}}, \bibinfo {author} {\bibfnamefont {K.}~\bibnamefont {Bennett}},
  \ and\ \bibinfo {author} {\bibfnamefont {S.}~\bibnamefont {Mukamel}},\ }\href
  {\doibase 10.1021/acs.jpclett.6b00864} {\bibfield  {journal} {\bibinfo
  {journal} {J. Phys. Chem. Lett.}\ }\textbf {\bibinfo {volume} {7}},\ \bibinfo
  {pages} {2050} (\bibinfo {year} {2016})}\BibitemShut {NoStop}%
\bibitem [{\citenamefont {Vendrell}(2018)}]{Vendrell2018Coherent}%
  \BibitemOpen
  \bibfield  {author} {\bibinfo {author} {\bibfnamefont {O.}~\bibnamefont
  {Vendrell}},\ }\href {\doibase 10.1016/j.chemphys.2018.02.008} {\bibfield
  {journal} {\bibinfo  {journal} {Chem. Phys.}\ }\textbf {\bibinfo {volume}
  {509}},\ \bibinfo {pages} {55} (\bibinfo {year} {2018})}\BibitemShut
  {NoStop}%
\bibitem [{\citenamefont {Triana}\ and\ \citenamefont
  {{Sanz-Vicario}}(2019)}]{Triana2019}%
  \BibitemOpen
  \bibfield  {author} {\bibinfo {author} {\bibfnamefont {J.~F.}\ \bibnamefont
  {Triana}}\ and\ \bibinfo {author} {\bibfnamefont {J.~L.}\ \bibnamefont
  {{Sanz-Vicario}}},\ }\href {\doibase 10.1103/PhysRevLett.122.063603}
  {\bibfield  {journal} {\bibinfo  {journal} {Phys. Rev. Lett.}\ }\textbf
  {\bibinfo {volume} {122}},\ \bibinfo {pages} {063603} (\bibinfo {year}
  {2019})}\BibitemShut {NoStop}%
\bibitem [{\citenamefont {Polli}\ \emph {et~al.}(2010)\citenamefont {Polli},
  \citenamefont {Alto{\`e}}, \citenamefont {Weingart}, \citenamefont
  {Spillane}, \citenamefont {Manzoni}, \citenamefont {Brida}, \citenamefont
  {Tomasello}, \citenamefont {Orlandi}, \citenamefont {Kukura}, \citenamefont
  {Mathies}, \citenamefont {Garavelli},\ and\ \citenamefont
  {Cerullo}}]{Polli2010}%
  \BibitemOpen
  \bibfield  {author} {\bibinfo {author} {\bibfnamefont {D.}~\bibnamefont
  {Polli}}, \bibinfo {author} {\bibfnamefont {P.}~\bibnamefont {Alto{\`e}}},
  \bibinfo {author} {\bibfnamefont {O.}~\bibnamefont {Weingart}}, \bibinfo
  {author} {\bibfnamefont {K.~M.}\ \bibnamefont {Spillane}}, \bibinfo {author}
  {\bibfnamefont {C.}~\bibnamefont {Manzoni}}, \bibinfo {author} {\bibfnamefont
  {D.}~\bibnamefont {Brida}}, \bibinfo {author} {\bibfnamefont
  {G.}~\bibnamefont {Tomasello}}, \bibinfo {author} {\bibfnamefont
  {G.}~\bibnamefont {Orlandi}}, \bibinfo {author} {\bibfnamefont
  {P.}~\bibnamefont {Kukura}}, \bibinfo {author} {\bibfnamefont {R.~A.}\
  \bibnamefont {Mathies}}, \bibinfo {author} {\bibfnamefont {M.}~\bibnamefont
  {Garavelli}}, \ and\ \bibinfo {author} {\bibfnamefont {G.}~\bibnamefont
  {Cerullo}},\ }\href {\doibase 10.1038/nature09346} {\bibfield  {journal}
  {\bibinfo  {journal} {Nature}\ }\textbf {\bibinfo {volume} {467}},\ \bibinfo
  {pages} {440} (\bibinfo {year} {2010})}\BibitemShut {NoStop}%
\bibitem [{\citenamefont {Berera}\ \emph {et~al.}(2009)\citenamefont {Berera},
  \citenamefont {van Grondelle},\ and\ \citenamefont {Kennis}}]{Berera2009}%
  \BibitemOpen
  \bibfield  {author} {\bibinfo {author} {\bibfnamefont {R.}~\bibnamefont
  {Berera}}, \bibinfo {author} {\bibfnamefont {R.}~\bibnamefont {van
  Grondelle}}, \ and\ \bibinfo {author} {\bibfnamefont {J.~T.~M.}\ \bibnamefont
  {Kennis}},\ }\href {\doibase 10.1007/s11120-009-9454-y} {\bibfield  {journal}
  {\bibinfo  {journal} {Photosynth. Res.}\ }\textbf {\bibinfo {volume} {101}},\
  \bibinfo {pages} {105} (\bibinfo {year} {2009})}\BibitemShut {NoStop}%
\bibitem [{\citenamefont {Feist}\ \emph {et~al.}(2018)\citenamefont {Feist},
  \citenamefont {Galego},\ and\ \citenamefont {{Garcia-Vidal}}}]{Feist2018}%
  \BibitemOpen
  \bibfield  {author} {\bibinfo {author} {\bibfnamefont {J.}~\bibnamefont
  {Feist}}, \bibinfo {author} {\bibfnamefont {J.}~\bibnamefont {Galego}}, \
  and\ \bibinfo {author} {\bibfnamefont {F.~J.}\ \bibnamefont
  {{Garcia-Vidal}}},\ }\href {\doibase 10.1021/acsphotonics.7b00680} {\bibfield
   {journal} {\bibinfo  {journal} {ACS Photonics}\ }\textbf {\bibinfo {volume}
  {5}},\ \bibinfo {pages} {205} (\bibinfo {year} {2018})}\BibitemShut {NoStop}%
\bibitem [{\citenamefont {Nicholson}\ \emph {et~al.}(1999)\citenamefont
  {Nicholson}, \citenamefont {Jasapara}, \citenamefont {Rudolph}, \citenamefont
  {Omenetto},\ and\ \citenamefont {Taylor}}]{Nicholson1999}%
  \BibitemOpen
  \bibfield  {author} {\bibinfo {author} {\bibfnamefont {J.~W.}\ \bibnamefont
  {Nicholson}}, \bibinfo {author} {\bibfnamefont {J.}~\bibnamefont {Jasapara}},
  \bibinfo {author} {\bibfnamefont {W.}~\bibnamefont {Rudolph}}, \bibinfo
  {author} {\bibfnamefont {F.~G.}\ \bibnamefont {Omenetto}}, \ and\ \bibinfo
  {author} {\bibfnamefont {A.~J.}\ \bibnamefont {Taylor}},\ }\href {\doibase
  10.1364/OL.24.001774} {\bibfield  {journal} {\bibinfo  {journal} {Opt. Lett.,
  OL}\ }\textbf {\bibinfo {volume} {24}},\ \bibinfo {pages} {1774} (\bibinfo
  {year} {1999})}\BibitemShut {NoStop}%
\bibitem [{\citenamefont {Iaconis}\ and\ \citenamefont
  {Walmsley}(1998)}]{Iaconis1998}%
  \BibitemOpen
  \bibfield  {author} {\bibinfo {author} {\bibfnamefont {C.}~\bibnamefont
  {Iaconis}}\ and\ \bibinfo {author} {\bibfnamefont {I.~A.}\ \bibnamefont
  {Walmsley}},\ }\href {\doibase 10.1364/OL.23.000792} {\bibfield  {journal}
  {\bibinfo  {journal} {Opt. Lett., OL}\ }\textbf {\bibinfo {volume} {23}},\
  \bibinfo {pages} {792} (\bibinfo {year} {1998})}\BibitemShut {NoStop}%
\bibitem [{\citenamefont {Trebino}\ \emph {et~al.}(1997)\citenamefont
  {Trebino}, \citenamefont {DeLong}, \citenamefont {Fittinghoff}, \citenamefont
  {Sweetser}, \citenamefont {Krumb{\"u}gel}, \citenamefont {Richman},\ and\
  \citenamefont {Kane}}]{Trebino1997}%
  \BibitemOpen
  \bibfield  {author} {\bibinfo {author} {\bibfnamefont {R.}~\bibnamefont
  {Trebino}}, \bibinfo {author} {\bibfnamefont {K.~W.}\ \bibnamefont {DeLong}},
  \bibinfo {author} {\bibfnamefont {D.~N.}\ \bibnamefont {Fittinghoff}},
  \bibinfo {author} {\bibfnamefont {J.~N.}\ \bibnamefont {Sweetser}}, \bibinfo
  {author} {\bibfnamefont {M.~A.}\ \bibnamefont {Krumb{\"u}gel}}, \bibinfo
  {author} {\bibfnamefont {B.~A.}\ \bibnamefont {Richman}}, \ and\ \bibinfo
  {author} {\bibfnamefont {D.~J.}\ \bibnamefont {Kane}},\ }\href {\doibase
  10.1063/1.1148286} {\bibfield  {journal} {\bibinfo  {journal} {Review of
  Scientific Instruments}\ }\textbf {\bibinfo {volume} {68}},\ \bibinfo {pages}
  {3277} (\bibinfo {year} {1997})}\BibitemShut {NoStop}%
\bibitem [{\citenamefont {Miranda}\ \emph {et~al.}(2012)\citenamefont
  {Miranda}, \citenamefont {Arnold}, \citenamefont {Fordell}, \citenamefont
  {Silva}, \citenamefont {Alonso}, \citenamefont {Weigand}, \citenamefont
  {L'Huillier},\ and\ \citenamefont {Crespo}}]{Miranda2012}%
  \BibitemOpen
  \bibfield  {author} {\bibinfo {author} {\bibfnamefont {M.}~\bibnamefont
  {Miranda}}, \bibinfo {author} {\bibfnamefont {C.~L.}\ \bibnamefont {Arnold}},
  \bibinfo {author} {\bibfnamefont {T.}~\bibnamefont {Fordell}}, \bibinfo
  {author} {\bibfnamefont {F.}~\bibnamefont {Silva}}, \bibinfo {author}
  {\bibfnamefont {B.}~\bibnamefont {Alonso}}, \bibinfo {author} {\bibfnamefont
  {R.}~\bibnamefont {Weigand}}, \bibinfo {author} {\bibfnamefont
  {A.}~\bibnamefont {L'Huillier}}, \ and\ \bibinfo {author} {\bibfnamefont
  {H.}~\bibnamefont {Crespo}},\ }\href {\doibase 10.1364/OE.20.018732}
  {\bibfield  {journal} {\bibinfo  {journal} {Opt. Express}\ }\textbf {\bibinfo
  {volume} {20}},\ \bibinfo {pages} {18732} (\bibinfo {year}
  {2012})}\BibitemShut {NoStop}%
\bibitem [{\citenamefont {Michetti}\ and\ \citenamefont
  {La~Rocca}(2009)}]{Michetti2009}%
  \BibitemOpen
  \bibfield  {author} {\bibinfo {author} {\bibfnamefont {P.}~\bibnamefont
  {Michetti}}\ and\ \bibinfo {author} {\bibfnamefont {G.~C.}\ \bibnamefont
  {La~Rocca}},\ }\href {\doibase 10.1103/PhysRevB.79.035325} {\bibfield
  {journal} {\bibinfo  {journal} {Phys. Rev. B}\ }\textbf {\bibinfo {volume}
  {79}},\ \bibinfo {pages} {035325} (\bibinfo {year} {2009})}\BibitemShut
  {NoStop}%
\bibitem [{\citenamefont {Kirton}\ and\ \citenamefont
  {Keeling}(2013)}]{Kirton2013}%
  \BibitemOpen
  \bibfield  {author} {\bibinfo {author} {\bibfnamefont {P.}~\bibnamefont
  {Kirton}}\ and\ \bibinfo {author} {\bibfnamefont {J.}~\bibnamefont
  {Keeling}},\ }\href {\doibase 10.1103/PhysRevLett.111.100404} {\bibfield
  {journal} {\bibinfo  {journal} {Phys. Rev. Lett.}\ }\textbf {\bibinfo
  {volume} {111}},\ \bibinfo {pages} {100404} (\bibinfo {year}
  {2013})}\BibitemShut {NoStop}%
\bibitem [{\citenamefont {Spano}(2015)}]{Spano2015}%
  \BibitemOpen
  \bibfield  {author} {\bibinfo {author} {\bibfnamefont {F.~C.}\ \bibnamefont
  {Spano}},\ }\href {\doibase 10.1063/1.4919348} {\bibfield  {journal}
  {\bibinfo  {journal} {J. Chem. Phys.}\ }\textbf {\bibinfo {volume} {142}},\
  \bibinfo {pages} {184707} (\bibinfo {year} {2015})}\BibitemShut {NoStop}%
\bibitem [{\citenamefont {Galego}\ \emph {et~al.}(2016)\citenamefont {Galego},
  \citenamefont {{Garcia-Vidal}},\ and\ \citenamefont {Feist}}]{Galego2016}%
  \BibitemOpen
  \bibfield  {author} {\bibinfo {author} {\bibfnamefont {J.}~\bibnamefont
  {Galego}}, \bibinfo {author} {\bibfnamefont {F.~J.}\ \bibnamefont
  {{Garcia-Vidal}}}, \ and\ \bibinfo {author} {\bibfnamefont {J.}~\bibnamefont
  {Feist}},\ }\href {\doibase 10.1038/ncomms13841} {\bibfield  {journal}
  {\bibinfo  {journal} {Nat. Commun.}\ }\textbf {\bibinfo {volume} {7}},\
  \bibinfo {pages} {13841} (\bibinfo {year} {2016})}\BibitemShut {NoStop}%
\bibitem [{\citenamefont {Galego}\ \emph {et~al.}(2017)\citenamefont {Galego},
  \citenamefont {{Garcia-Vidal}},\ and\ \citenamefont {Feist}}]{Galego2017}%
  \BibitemOpen
  \bibfield  {author} {\bibinfo {author} {\bibfnamefont {J.}~\bibnamefont
  {Galego}}, \bibinfo {author} {\bibfnamefont {F.~J.}\ \bibnamefont
  {{Garcia-Vidal}}}, \ and\ \bibinfo {author} {\bibfnamefont {J.}~\bibnamefont
  {Feist}},\ }\href {\doibase 10.1103/PhysRevLett.119.136001} {\bibfield
  {journal} {\bibinfo  {journal} {Phys. Rev. Lett.}\ }\textbf {\bibinfo
  {volume} {119}},\ \bibinfo {pages} {136001} (\bibinfo {year}
  {2017})}\BibitemShut {NoStop}%
\bibitem [{\citenamefont {Schr{\"o}der}\ \emph {et~al.}(2019)\citenamefont
  {Schr{\"o}der}, \citenamefont {Turban}, \citenamefont {Musser}, \citenamefont
  {Hine},\ and\ \citenamefont {Chin}}]{Schroder2019}%
  \BibitemOpen
  \bibfield  {author} {\bibinfo {author} {\bibfnamefont {F.~A. Y.~N.}\
  \bibnamefont {Schr{\"o}der}}, \bibinfo {author} {\bibfnamefont {D.~H.~P.}\
  \bibnamefont {Turban}}, \bibinfo {author} {\bibfnamefont {A.~J.}\
  \bibnamefont {Musser}}, \bibinfo {author} {\bibfnamefont {N.~D.~M.}\
  \bibnamefont {Hine}}, \ and\ \bibinfo {author} {\bibfnamefont {A.~W.}\
  \bibnamefont {Chin}},\ }\href {\doibase 10.1038/s41467-019-09039-7}
  {\bibfield  {journal} {\bibinfo  {journal} {Nat. Commun.}\ }\textbf {\bibinfo
  {volume} {10}},\ \bibinfo {pages} {1062} (\bibinfo {year}
  {2019})}\BibitemShut {NoStop}%
\bibitem [{\citenamefont {{K{\'e}na-Cohen}}\ and\ \citenamefont
  {Forrest}(2010)}]{Kena-Cohen2010}%
  \BibitemOpen
  \bibfield  {author} {\bibinfo {author} {\bibfnamefont {S.}~\bibnamefont
  {{K{\'e}na-Cohen}}}\ and\ \bibinfo {author} {\bibfnamefont {S.~R.}\
  \bibnamefont {Forrest}},\ }\href {\doibase 10.1038/nphoton.2010.86}
  {\bibfield  {journal} {\bibinfo  {journal} {Nat. Photonics}\ }\textbf
  {\bibinfo {volume} {4}},\ \bibinfo {pages} {371} (\bibinfo {year}
  {2010})}\BibitemShut {NoStop}%
\bibitem [{\citenamefont {Ramezani}\ \emph {et~al.}(2017)\citenamefont
  {Ramezani}, \citenamefont {Halpin}, \citenamefont
  {{Fern{\'a}ndez-Dom{\'i}nguez}}, \citenamefont {Feist}, \citenamefont
  {Rodriguez}, \citenamefont {{Garcia-Vidal}},\ and\ \citenamefont
  {G{\'o}mez~Rivas}}]{Ramezani2017Plasmon}%
  \BibitemOpen
  \bibfield  {author} {\bibinfo {author} {\bibfnamefont {M.}~\bibnamefont
  {Ramezani}}, \bibinfo {author} {\bibfnamefont {A.}~\bibnamefont {Halpin}},
  \bibinfo {author} {\bibfnamefont {A.~I.}\ \bibnamefont
  {{Fern{\'a}ndez-Dom{\'i}nguez}}}, \bibinfo {author} {\bibfnamefont
  {J.}~\bibnamefont {Feist}}, \bibinfo {author} {\bibfnamefont {S.~R.-K.}\
  \bibnamefont {Rodriguez}}, \bibinfo {author} {\bibfnamefont {F.~J.}\
  \bibnamefont {{Garcia-Vidal}}}, \ and\ \bibinfo {author} {\bibfnamefont
  {J.}~\bibnamefont {G{\'o}mez~Rivas}},\ }\href {\doibase
  10.1364/OPTICA.4.000031} {\bibfield  {journal} {\bibinfo  {journal} {Optica}\
  }\textbf {\bibinfo {volume} {4}},\ \bibinfo {pages} {31} (\bibinfo {year}
  {2017})}\BibitemShut {NoStop}%
\bibitem [{\citenamefont {Acuna}\ \emph {et~al.}(2012)\citenamefont {Acuna},
  \citenamefont {M{\"o}ller}, \citenamefont {Holzmeister}, \citenamefont
  {Beater}, \citenamefont {Lalkens},\ and\ \citenamefont
  {Tinnefeld}}]{Acuna2012}%
  \BibitemOpen
  \bibfield  {author} {\bibinfo {author} {\bibfnamefont {G.~P.}\ \bibnamefont
  {Acuna}}, \bibinfo {author} {\bibfnamefont {F.~M.}\ \bibnamefont
  {M{\"o}ller}}, \bibinfo {author} {\bibfnamefont {P.}~\bibnamefont
  {Holzmeister}}, \bibinfo {author} {\bibfnamefont {S.}~\bibnamefont {Beater}},
  \bibinfo {author} {\bibfnamefont {B.}~\bibnamefont {Lalkens}}, \ and\
  \bibinfo {author} {\bibfnamefont {P.}~\bibnamefont {Tinnefeld}},\ }\href
  {\doibase 10.1126/science.1228638} {\bibfield  {journal} {\bibinfo  {journal}
  {Science}\ }\textbf {\bibinfo {volume} {338}},\ \bibinfo {pages} {506}
  (\bibinfo {year} {2012})}\BibitemShut {NoStop}%
\bibitem [{\citenamefont {Akselrod}\ \emph {et~al.}(2014)\citenamefont
  {Akselrod}, \citenamefont {Argyropoulos}, \citenamefont {Hoang},
  \citenamefont {Cirac{\`i}}, \citenamefont {Fang}, \citenamefont {Huang},
  \citenamefont {Smith},\ and\ \citenamefont
  {Mikkelsen}}]{Akselrod2014Probing}%
  \BibitemOpen
  \bibfield  {author} {\bibinfo {author} {\bibfnamefont {G.~M.}\ \bibnamefont
  {Akselrod}}, \bibinfo {author} {\bibfnamefont {C.}~\bibnamefont
  {Argyropoulos}}, \bibinfo {author} {\bibfnamefont {T.~B.}\ \bibnamefont
  {Hoang}}, \bibinfo {author} {\bibfnamefont {C.}~\bibnamefont {Cirac{\`i}}},
  \bibinfo {author} {\bibfnamefont {C.}~\bibnamefont {Fang}}, \bibinfo {author}
  {\bibfnamefont {J.}~\bibnamefont {Huang}}, \bibinfo {author} {\bibfnamefont
  {D.~R.}\ \bibnamefont {Smith}}, \ and\ \bibinfo {author} {\bibfnamefont
  {M.~H.}\ \bibnamefont {Mikkelsen}},\ }\href {\doibase
  10.1038/nphoton.2014.228} {\bibfield  {journal} {\bibinfo  {journal} {Nature
  Photon}\ }\textbf {\bibinfo {volume} {8}},\ \bibinfo {pages} {835} (\bibinfo
  {year} {2014})}\BibitemShut {NoStop}%
\bibitem [{\citenamefont {Kongsuwan}\ \emph {et~al.}(2018)\citenamefont
  {Kongsuwan}, \citenamefont {Demetriadou}, \citenamefont {Chikkaraddy},
  \citenamefont {Benz}, \citenamefont {Turek}, \citenamefont {Keyser},
  \citenamefont {Baumberg},\ and\ \citenamefont {Hess}}]{Kongsuwan2018}%
  \BibitemOpen
  \bibfield  {author} {\bibinfo {author} {\bibfnamefont {N.}~\bibnamefont
  {Kongsuwan}}, \bibinfo {author} {\bibfnamefont {A.}~\bibnamefont
  {Demetriadou}}, \bibinfo {author} {\bibfnamefont {R.}~\bibnamefont
  {Chikkaraddy}}, \bibinfo {author} {\bibfnamefont {F.}~\bibnamefont {Benz}},
  \bibinfo {author} {\bibfnamefont {V.~A.}\ \bibnamefont {Turek}}, \bibinfo
  {author} {\bibfnamefont {U.~F.}\ \bibnamefont {Keyser}}, \bibinfo {author}
  {\bibfnamefont {J.~J.}\ \bibnamefont {Baumberg}}, \ and\ \bibinfo {author}
  {\bibfnamefont {O.}~\bibnamefont {Hess}},\ }\href {\doibase
  10.1021/acsphotonics.7b00668} {\bibfield  {journal} {\bibinfo  {journal} {ACS
  Photonics}\ }\textbf {\bibinfo {volume} {5}},\ \bibinfo {pages} {186}
  (\bibinfo {year} {2018})}\BibitemShut {NoStop}%
\bibitem [{\citenamefont {Baumberg}\ \emph {et~al.}(2019)\citenamefont
  {Baumberg}, \citenamefont {Aizpurua}, \citenamefont {Mikkelsen},\ and\
  \citenamefont {Smith}}]{Baumberg2019}%
  \BibitemOpen
  \bibfield  {author} {\bibinfo {author} {\bibfnamefont {J.~J.}\ \bibnamefont
  {Baumberg}}, \bibinfo {author} {\bibfnamefont {J.}~\bibnamefont {Aizpurua}},
  \bibinfo {author} {\bibfnamefont {M.~H.}\ \bibnamefont {Mikkelsen}}, \ and\
  \bibinfo {author} {\bibfnamefont {D.~R.}\ \bibnamefont {Smith}},\ }\href
  {\doibase 10.1038/s41563-019-0290-y} {\bibfield  {journal} {\bibinfo
  {journal} {Nat. Mater.}\ }\textbf {\bibinfo {volume} {18}},\ \bibinfo {pages}
  {668} (\bibinfo {year} {2019})}\BibitemShut {NoStop}%
\bibitem [{\citenamefont {Frisch}\ \emph {et~al.}(2009)\citenamefont {Frisch},
  \citenamefont {Trucks}, \citenamefont {Schlegel}, \citenamefont {Scuseria},
  \citenamefont {Robb}, \citenamefont {Cheeseman}, \citenamefont {Scalmani},
  \citenamefont {Barone}, \citenamefont {Petersson}, \citenamefont {Nakatsuji},
  \citenamefont {Li}, \citenamefont {Caricato}, \citenamefont {Marenich},
  \citenamefont {Bloino}, \citenamefont {Janesko}, \citenamefont {Gomperts},
  \citenamefont {Mennucci}, \citenamefont {Hratchian}, \citenamefont {Ortiz},
  \citenamefont {Izmaylov}, \citenamefont {Sonnenberg}, \citenamefont
  {{Williams-Young}}, \citenamefont {Ding}, \citenamefont {Lipparini},
  \citenamefont {Egidi}, \citenamefont {Goings}, \citenamefont {Peng},
  \citenamefont {Petrone}, \citenamefont {Henderson}, \citenamefont
  {Ranasinghe}, \citenamefont {Zakrzewski}, \citenamefont {Gao}, \citenamefont
  {Rega}, \citenamefont {Zheng}, \citenamefont {Liang}, \citenamefont {Hada},
  \citenamefont {Ehara}, \citenamefont {Toyota}, \citenamefont {Fukuda},
  \citenamefont {Hasegawa}, \citenamefont {Ishida}, \citenamefont {Nakajima},
  \citenamefont {Honda}, \citenamefont {Kitao}, \citenamefont {Nakai},
  \citenamefont {Vreven}, \citenamefont {Throssell}, \citenamefont
  {Montgomery}, \citenamefont {{Jr.,}}, \citenamefont {Peralta}, \citenamefont
  {Ogliaro}, \citenamefont {Bearpark}, \citenamefont {Heyd}, \citenamefont
  {Brothers}, \citenamefont {Kudin}, \citenamefont {Staroverov}, \citenamefont
  {Keith}, \citenamefont {Kobayashi}, \citenamefont {Normand}, \citenamefont
  {Raghavachari}, \citenamefont {Rendell}, \citenamefont {Burant},
  \citenamefont {Iyengar}, \citenamefont {Tomasi}, \citenamefont {Cossi},
  \citenamefont {Millam}, \citenamefont {Klene}, \citenamefont {Adamo},
  \citenamefont {Cammi}, \citenamefont {Ochterski}, \citenamefont {Martin},
  \citenamefont {Morokuma}, \citenamefont {Farkas}, \citenamefont {Foresman},\
  and\ \citenamefont {Fox}}]{Gaussian09E01}%
  \BibitemOpen
  \bibfield  {author} {\bibinfo {author} {\bibfnamefont {M.~J.}\ \bibnamefont
  {Frisch}}, \bibinfo {author} {\bibfnamefont {G.~W.}\ \bibnamefont {Trucks}},
  \bibinfo {author} {\bibfnamefont {H.~B.}\ \bibnamefont {Schlegel}}, \bibinfo
  {author} {\bibfnamefont {G.~E.}\ \bibnamefont {Scuseria}}, \bibinfo {author}
  {\bibfnamefont {M.~A.}\ \bibnamefont {Robb}}, \bibinfo {author}
  {\bibfnamefont {J.~R.}\ \bibnamefont {Cheeseman}}, \bibinfo {author}
  {\bibfnamefont {G.}~\bibnamefont {Scalmani}}, \bibinfo {author}
  {\bibfnamefont {V.}~\bibnamefont {Barone}}, \bibinfo {author} {\bibfnamefont
  {G.~A.}\ \bibnamefont {Petersson}}, \bibinfo {author} {\bibfnamefont
  {H.}~\bibnamefont {Nakatsuji}}, \bibinfo {author} {\bibfnamefont
  {X.}~\bibnamefont {Li}}, \bibinfo {author} {\bibfnamefont {M.}~\bibnamefont
  {Caricato}}, \bibinfo {author} {\bibfnamefont {A.}~\bibnamefont {Marenich}},
  \bibinfo {author} {\bibfnamefont {J.}~\bibnamefont {Bloino}}, \bibinfo
  {author} {\bibfnamefont {B.~G.}\ \bibnamefont {Janesko}}, \bibinfo {author}
  {\bibfnamefont {R.}~\bibnamefont {Gomperts}}, \bibinfo {author}
  {\bibfnamefont {B.}~\bibnamefont {Mennucci}}, \bibinfo {author}
  {\bibfnamefont {H.~P.}\ \bibnamefont {Hratchian}}, \bibinfo {author}
  {\bibfnamefont {J.~V.}\ \bibnamefont {Ortiz}}, \bibinfo {author}
  {\bibfnamefont {A.~F.}\ \bibnamefont {Izmaylov}}, \bibinfo {author}
  {\bibfnamefont {J.~L.}\ \bibnamefont {Sonnenberg}}, \bibinfo {author}
  {\bibfnamefont {D.}~\bibnamefont {{Williams-Young}}}, \bibinfo {author}
  {\bibfnamefont {F.}~\bibnamefont {Ding}}, \bibinfo {author} {\bibfnamefont
  {F.}~\bibnamefont {Lipparini}}, \bibinfo {author} {\bibfnamefont
  {F.}~\bibnamefont {Egidi}}, \bibinfo {author} {\bibfnamefont
  {J.}~\bibnamefont {Goings}}, \bibinfo {author} {\bibfnamefont
  {B.}~\bibnamefont {Peng}}, \bibinfo {author} {\bibfnamefont {A.}~\bibnamefont
  {Petrone}}, \bibinfo {author} {\bibfnamefont {T.}~\bibnamefont {Henderson}},
  \bibinfo {author} {\bibfnamefont {D.}~\bibnamefont {Ranasinghe}}, \bibinfo
  {author} {\bibfnamefont {V.~G.}\ \bibnamefont {Zakrzewski}}, \bibinfo
  {author} {\bibfnamefont {J.}~\bibnamefont {Gao}}, \bibinfo {author}
  {\bibfnamefont {N.}~\bibnamefont {Rega}}, \bibinfo {author} {\bibfnamefont
  {G.}~\bibnamefont {Zheng}}, \bibinfo {author} {\bibfnamefont
  {W.}~\bibnamefont {Liang}}, \bibinfo {author} {\bibfnamefont
  {M.}~\bibnamefont {Hada}}, \bibinfo {author} {\bibfnamefont {M.}~\bibnamefont
  {Ehara}}, \bibinfo {author} {\bibfnamefont {K.}~\bibnamefont {Toyota}},
  \bibinfo {author} {\bibfnamefont {R.}~\bibnamefont {Fukuda}}, \bibinfo
  {author} {\bibfnamefont {J.}~\bibnamefont {Hasegawa}}, \bibinfo {author}
  {\bibfnamefont {M.}~\bibnamefont {Ishida}}, \bibinfo {author} {\bibfnamefont
  {T.}~\bibnamefont {Nakajima}}, \bibinfo {author} {\bibfnamefont
  {Y.}~\bibnamefont {Honda}}, \bibinfo {author} {\bibfnamefont
  {O.}~\bibnamefont {Kitao}}, \bibinfo {author} {\bibfnamefont
  {H.}~\bibnamefont {Nakai}}, \bibinfo {author} {\bibfnamefont
  {T.}~\bibnamefont {Vreven}}, \bibinfo {author} {\bibfnamefont
  {K.}~\bibnamefont {Throssell}}, \bibinfo {author} {\bibfnamefont {J.~A.}\
  \bibnamefont {Montgomery}}, \bibinfo {author} {\bibnamefont {{Jr.,}}},
  \bibinfo {author} {\bibfnamefont {J.~E.}\ \bibnamefont {Peralta}}, \bibinfo
  {author} {\bibfnamefont {F.}~\bibnamefont {Ogliaro}}, \bibinfo {author}
  {\bibfnamefont {M.}~\bibnamefont {Bearpark}}, \bibinfo {author}
  {\bibfnamefont {J.~J.}\ \bibnamefont {Heyd}}, \bibinfo {author}
  {\bibfnamefont {E.}~\bibnamefont {Brothers}}, \bibinfo {author}
  {\bibfnamefont {K.~N.}\ \bibnamefont {Kudin}}, \bibinfo {author}
  {\bibfnamefont {V.~N.}\ \bibnamefont {Staroverov}}, \bibinfo {author}
  {\bibfnamefont {T.}~\bibnamefont {Keith}}, \bibinfo {author} {\bibfnamefont
  {R.}~\bibnamefont {Kobayashi}}, \bibinfo {author} {\bibfnamefont
  {J.}~\bibnamefont {Normand}}, \bibinfo {author} {\bibfnamefont
  {K.}~\bibnamefont {Raghavachari}}, \bibinfo {author} {\bibfnamefont
  {A.}~\bibnamefont {Rendell}}, \bibinfo {author} {\bibfnamefont {J.~C.}\
  \bibnamefont {Burant}}, \bibinfo {author} {\bibfnamefont {S.~S.}\
  \bibnamefont {Iyengar}}, \bibinfo {author} {\bibfnamefont {J.}~\bibnamefont
  {Tomasi}}, \bibinfo {author} {\bibfnamefont {M.}~\bibnamefont {Cossi}},
  \bibinfo {author} {\bibfnamefont {J.~M.}\ \bibnamefont {Millam}}, \bibinfo
  {author} {\bibfnamefont {M.}~\bibnamefont {Klene}}, \bibinfo {author}
  {\bibfnamefont {C.}~\bibnamefont {Adamo}}, \bibinfo {author} {\bibfnamefont
  {R.}~\bibnamefont {Cammi}}, \bibinfo {author} {\bibfnamefont {J.~W.}\
  \bibnamefont {Ochterski}}, \bibinfo {author} {\bibfnamefont {R.~L.}\
  \bibnamefont {Martin}}, \bibinfo {author} {\bibfnamefont {K.}~\bibnamefont
  {Morokuma}}, \bibinfo {author} {\bibfnamefont {O.}~\bibnamefont {Farkas}},
  \bibinfo {author} {\bibfnamefont {J.~B.}\ \bibnamefont {Foresman}}, \ and\
  \bibinfo {author} {\bibfnamefont {D.~J.}\ \bibnamefont {Fox}},\ }\href@noop
  {} {\enquote {\bibinfo {title} {{Gaussian 09, {{Revision E}}.01}},}\
  }\bibinfo {howpublished} {Gaussian, Inc} (\bibinfo {year} {2009})\BibitemShut
  {NoStop}%
\bibitem [{\citenamefont {Duschinsky}(1937)}]{Duschinsky1937}%
  \BibitemOpen
  \bibfield  {author} {\bibinfo {author} {\bibfnamefont {F.}~\bibnamefont
  {Duschinsky}},\ }\href@noop {} {\bibfield  {journal} {\bibinfo  {journal}
  {Acta Physicochim URSS}\ }\textbf {\bibinfo {volume} {7}},\ \bibinfo {pages}
  {551} (\bibinfo {year} {1937})}\BibitemShut {NoStop}%
\bibitem [{\citenamefont {Chin}\ \emph {et~al.}(2010)\citenamefont {Chin},
  \citenamefont {Rivas}, \citenamefont {Huelga},\ and\ \citenamefont
  {Plenio}}]{Chin2010}%
  \BibitemOpen
  \bibfield  {author} {\bibinfo {author} {\bibfnamefont {A.~W.}\ \bibnamefont
  {Chin}}, \bibinfo {author} {\bibfnamefont {{\'A}.}~\bibnamefont {Rivas}},
  \bibinfo {author} {\bibfnamefont {S.~F.}\ \bibnamefont {Huelga}}, \ and\
  \bibinfo {author} {\bibfnamefont {M.~B.}\ \bibnamefont {Plenio}},\ }\href
  {\doibase 10.1063/1.3490188} {\bibfield  {journal} {\bibinfo  {journal} {J.
  Math. Phys.}\ }\textbf {\bibinfo {volume} {51}},\ \bibinfo {pages} {092109}
  (\bibinfo {year} {2010})}\BibitemShut {NoStop}%
\end{thebibliography}%

\end{document}